\documentclass[letterpaper]{JHEP3}
\usepackage{amsmath, amsfonts, xspace}
\usepackage{epsfig}
\usepackage{cite}
\usepackage{array}
\usepackage{yfonts}
\usepackage{cite}

\def\scrO{\mathcal{O}}

\def\F{f}
\def\Fij{F_{ij}}
\def\G{g}
\def\Gij{G_{ij}}
\def\Ginv{G^{ij}}
\def\Gcan{g_{\rm can}}
\def\Fcan{f_{\rm can}}
\def\Fd{{\mathcal F}}
\def\can{{\rm can}}
\def\lap{\triangle}
\def\ls{\Delta}

\def\Mm{M}

\def\be{\begin{equation}}
\def\ee{\end{equation}}
\def\Vol{\mbox{Vol}}

\newcommand{\Z}{\mathbb{Z}}
\newcommand{\R}{\mathbb{R}}

\newcommand{\pd}[2]{\frac{\partial #1}{\partial #2}}
\newcommand{\K}{K\"ahler\xspace}
\newcommand{\CP}{\mathbb{CP}}

\DeclareMathOperator{\sech}{sech}

\preprint{{\tt hep-th/0703057} \\
SU-ITP-07/01\\
Imperial/TP/07/TW/01\\
NSF-KITP-07-32
}

\title{Numerical \K-Einstein metric on the third del Pezzo}

\author{Charles Doran \\ Department of Mathematics, University of Washington, Seattle, WA 98195-1560, USA \\ \email{doran@math.washington.edu} }

\author{Matthew Headrick \\ Stanford Institute for Theoretical Physics, Stanford, CA 94305-4060, USA \\ \email{headrick@stanford.edu}}

\author{Christopher P. Herzog \\ Department of Physics, University of Washington, Seattle, WA 98195-1560, USA \\ \email{herzog@phys.washington.edu} }

\author{Joshua Kantor \\ Department of Mathematics, University of Washington, Seattle, WA 98195-1560, USA \\ \email{jkantor@math.washington.edu} }

\author{Toby Wiseman \\ Blackett Laboratory, Imperial College London, London SW7 2AZ, UK \\ \email{t.wiseman@ic.ac.uk} }

\abstract{
The third del Pezzo surface admits a unique \K-Einstein metric, which is not known in closed form. The manifold's toric structure reduces the Einstein equation to a single Monge-Amp\`ere equation in two real dimensions. We numerically solve this nonlinear PDE using three different algorithms, and describe the resulting metric. The first two algorithms involve simulation of Ricci flow, in complex and symplectic coordinates respectively. The third algorithm involves turning the PDE into an optimization problem on a certain space of metrics, which are symplectic analogues of the ``algebraic" metrics used in numerical work on Calabi-Yau manifolds. Our algorithms should be applicable to general toric %Fano 
manifolds. Using our metric, we compute various geometric quantities of interest, including Laplacian eigenvalues and a harmonic $(1,1)$-form. The metric and $(1,1)$-form can be used to construct a Klebanov-Tseytlin-like supergravity solution.
}

\begin{document}

\section{Introduction}

\K metrics on manifolds play an important role in mathematics and physics.
As Yau demonstrated \cite{Yau:1977ms}, in the \K case it is often possible to
prove the existence (or nonexistence) of metrics which solve the Einstein equation.
While it is extremely valuable to know whether they exist, for many purposes one also wants to know their specific form. The existence theorems, however, are generally non-constructive, and explicit examples of \K-Einstein metrics are rare. This state of affairs naturally leads to the following question: Is it possible, in practice, to find accurate numerical approximations to these metrics using computers? The last two years have seen significant success, with a variety of different algorithms providing numerical solutions to the Einstein equation on the Calabi-Yau surface K3 \cite{HW,Donaldson,Douglasetal1} and on a three-fold \cite{Douglasetal2}. The \K property proved to be as crucial for this numerical work as it was for the existence theorems.

In this paper we extend this success to a non-Calabi-Yau manifold, namely $\CP^2$ blown up at three points ($\CP^2\#3\overline{\CP}^2$), also known as the third del Pezzo surface ($dP_3$). This manifold is known by work of Siu \cite{Siu} and Tian-Yau \cite{TianYau} to admit a \K-Einstein metric with positive cosmological constant, but as in the Calabi-Yau case that metric is not known explicitly. An important property of $dP_3$ that differentiates it from Calabi-Yau manifolds is that it is \emph{toric}. Toric manifolds are a special class of \K manifolds whose $U(1)^n$ isometry group (where $n$ is the manifold's complex dimension) allows even greater analytical control, and we develop algorithms for solving the Einstein equation that specifically exploit this structure. We should also note that the \K-Einstein metric on $dP_3$ is unique (up to rescaling); therefore, rather than having a moduli space of metrics as we have in the case of Calabi-Yau manifolds, there is only one metric to compute.

On the physics side, the \K-Einstein metric on $dP_3$ is important because it can be used to construct an example of a gauge/gravity duality. These dualities provide a bridge between physical theories of radically different character, allowing computation in one theory using the methods of its dual. Given the \K-Einstein metric on $dP_3$, one can construct a five-dimensional Sasaki-Einstein metric. Compactifying type IIB supergravity on this manifold one obtains an AdS${}_5$ supergravity solution, which has a known superconformal gauge theory dual \cite{Feng:2001xr,Hanany:2001py,Beasley:2001zp}. An interesting generalization includes a 3-form flux from wrapped D5-branes. This flux can be written in terms of a harmonic $(1,1)$-form on $dP_3$, which we also compute numerically in this paper. The resulting supergravity solution (the analogue of the Klebanov-Tseytlin solution on the conifold \cite{KT}) is nakedly singular, but is the first step toward finding the full supergravity solution on the smoothed-out cone (the analogue of the Klebanov-Strassler warped deformed conifold \cite{KS}). This smoothed-out cone is dual to a cascading gauge theory, and knowing the explicit form of the supergravity solution would be useful for both gauge theory and cosmology applications. This physics background is explained in detail in Section \ref{sec:physics}.

In Section \ref{sec:mathbackground}, we review the mathematical background necessary for understanding the rest of the paper. Here we closely follow the review article on toric geometry by Abreu \cite{Abreu}. We explain the two natural coordinate systems on a toric manifold, namely complex and symplectic coordinates. In complex coordinates, the metric is encoded in the \K potential, and in symplectic coordinates in the symplectic potential; these two functions are related by a Legendre transform. In either coordinate system, the Einstein equation reduces to a single nonlinear partial differential equation, of Monge-Amp\`ere type, for the corresponding potential. Thanks to the $U(1)^n$ symmetry, this PDE is in half the number of dimensions of the original manifold (two real dimensions for $dP_3$). In this section we also derive the equation we need for the $(1,1)$-form, and give all the necessary details about $dP_3$.

In this paper we describe three different methods to solve the Monge-Amp\`ere equation. In Section \ref{sec:ricciflow} we explain the first two methods, which involve numerically simulating Ricci flow in complex and symplectic coordinates respectively. Specifically, we use a variant of Ricci flow (normalized) that includes a $\Lambda$ term,
\be\label{RFdef1}
\pd{g_{\mu\nu}}{t} = -2R_{\mu\nu}+2\Lambda g_{\mu\nu},
\ee
whose fixed points are clearly Einstein metrics with cosmological constant $\Lambda$. According to a recent result of Tian-Zhu \cite{Tian}, on a manifold which admits a \K-Einstein metric, the flow \eqref{RFdef} converges to it starting from any metric in the same \K class. Our simulations behaved accordingly, yielding \K-Einstein metrics %which agreed with each other to within and whose independent
accurate (at the highest resolutions we employed) to a few parts in $10^6$ (and which agree with each other to within that error). 
%and whose independent accuracy we estimate to be of the same order of magnitude.
%at around one part in $10^6$
%, and which agreed with each other to that accuracy. 
Numerical simulations of Ricci flow have been studied before in a variety of contexts \cite{Hori:2001ax,Garfinkle,Headrick:2006ti}, but as far as we know this is the first time they have been used to find a new solution to the Einstein equation (aside from a limited exploration of its use on K3 \cite{HW}). In this section we also explore the geometry of this solution, and discuss a method for computing eigenvalues and eigenfunctions of the Laplacian in that background.

In Section \ref{sec:polynomial} we introduce a different way to represent the metric, based on certain polynomials in the symplectic coordinates. This non-local representation is a symplectic analogue of the ``algebraic" metrics on Calabi-Yau manifolds employed in numerical work by Donaldson \cite{Donaldson} and Douglas et al.\ \cite{Douglasetal1,Douglasetal2}. To demonstrate the utility of this representation we give a low order polynomial fit to the numerical solutions found by Ricci flow, that can be written on one line and yet agrees with the true solution to one part in $10^3$, and everywhere satisfies the Einstein condition to better than $10\%$.

In Section \ref{sec:leastsquares} we discuss our third method to compute the \K-Einstein metric. As above we represent the metric using polynomials in the symplectic coordinates, but now we constrain the polynomial coefficients by solving the Monge-Amp\`ere equation order by order in the coordinates. This leaves a small number of undetermined coefficients, which we compute by minimizing an error function. Using this method we obtain numerical metrics of similar accuracy to those found by Ricci flow. By a similar method we also calculate eigenvalues and eigenfunctions of the Laplacian, and the harmonic $(1,1)$-form.  

We conclude the paper with a brief discussion of the three methods and their relative merits in Section \ref{sec:discussion}.

At the two websites \cite{website}, we have made available for download the full numerical data representing our metrics, as well as Mathematica notebooks that input the data and allow the user to work with those metrics. The codes used to generate the data are also available on those websites.

We believe that all of the methods we present here can be applied to a general toric manifold. We will report elsewhere on an application to $dP_2$ \cite{HWmaybe}, which does not admit a \K-Einstein metric but does admit a \K-Ricci soliton \cite{WA}. For future work, there is also a natural analogue of $dP_3$ to study in three complex dimensions.  From Batyrev's classification of toric Fano threefolds, it follows that there are precisely two of them that admit \K-Einstein metrics which are not themselves products of lower dimensional manifolds \cite[Section 4]{BA}. One of these is $\mathbb{CP}^3$.  The other is the total space of the projectivization of the rank two bundle $\scrO \oplus \scrO(1,-1)$ over $\mathbb{CP}^1 \times \mathbb{CP}^1$. Its Delzant polytope (see Section \ref{sec:mathbackground} for the definition) possesses a $D_4$ symmetry, and a fundamental region is simply a tetrahedron.

While this work was in progress we have learned that \K-Einstein metrics on $dP_3$ have also been computed in \cite{DonaldsonKeller} using the methods of \cite{Donaldson}.

\section{Gauge/gravity duality}
\label{sec:physics}

A prototypical example of a gauge/gravity duality which provides the physics motivation
for studying $dP_3$ is the Klebanov-Strassler (KS) 
supergravity solution \cite{KS}.  (Mathematicians may wish to skip this section.) The KS solution is a solution of the type IIB
supergravity equations of motion.  The space-time is a warped product
of Minkowski space $\mathbb{R}^{1,3}$ and the deformed conifold $X$.
The affine variety $X$
 has an embedding in $\mathbb{C}^4$ defined by
 \be
\sum_{i=1}^4 z_i^2 = \epsilon
\ee
where $z_i \in \mathbb{C}$.  There are also a variety of nontrivial 
fluxes in this solution which we will return to later.  

One important aspect of the KS solution is its conjectured duality to a
non-abelian gauge theory, namely the cascading $SU(N) \times SU(N+M)$ 
${\mathcal N}=1$ supersymmetric gauge theory with bifundamental fields
$A_i$ and $B_i$, $i=1$ or 2, and superpotential
\be
W = \epsilon_{ij} \epsilon_{kl} A_i B_k A_j B_l \ .
\ee
This theory is similar in a number of respects to QCD; it exhibits 
renormalization group flow, chiral symmetry breaking, and confinement.
Moreover, all of these properties can be understood from the dual
gravitational perspective.  

In addition to its gauge theory applications, the KS solution is important for
cosmology.  Treating the deformed conifold as a local feature
of a compact Calabi-Yau manifold, the KS solution provides a string
compactification with a natural hierarchy of scales
in which all the complex moduli are fixed \cite{GKPflux}.  
Stabilizing the \K moduli as well \cite{KKLT}, the KS solution can become a
 metastable string vacuum and thus a model of the real world. 
In this context, inflation might correspond to the motion of 
D-branes \cite{Kachru:2003sx} and cosmic strings might be the fundamental
and D-strings of type IIB string theory \cite{Copeland:2003bj}.
 
One naturally wonders to what extent the cosmological and gauge theoretic
applications depend upon the choice of the deformed conifold.  
A natural way to generalize $X$ is to consider smoothings of other 
Calabi-Yau singularities. 
One such family of singularities involves
a Calabi-Yau where a del Pezzo surface $dP_n$ shrinks to zero size. 
(Here $dP_n$ is $\mathbb{CP}^2$ blown up at $n$ points.)
Note we are distinguishing here between resolutions --- or \K structure
deformations --- where even dimensional cycles are made to be of finite size,
and smoothings --- complex structure deformations --- where a three
dimensional cycle is made finite.
Using toric geometry techniques, Altmann \cite{Altmann} has shown that the total space of the canonical
bundle over $dP_1$ admits no smoothings, while $dP_2$ admits one and
$dP_3$ two.\footnote{%
The physical relevance of this fact for supersymmetry breaking was pointed out in
\cite{BHOP, Franco, Bertolini} (see also \cite{Forcella} for a more recent account).
}
   The higher $dP_n$ are not toric.
Thus two relatively simple
candidates for generalizing the KS solution are smoothed cones over
the complex surfaces $dP_2$ and $dP_3$.

Without knowing the details of the metric on the smoothed cone $X$, one can show that
a generalization of the KS solution exists for such warped products
\cite{GranaPolchinski}.  The solution
will have a ten dimensional line element of the form
\be
ds^2 = h(p)^{-1/2} \eta_{\mu\nu} dx^\mu dx^\nu  + h(p)^{1/2} ds_X^2 \ ,
\ee
where $ds_X^2$ is the line element on $X$, $p \in X$, the map $h: X \rightarrow \mathbb{R}^+$ is
called the ``warp factor'', and $\eta_{\mu\nu}$ is the Minkowski tensor for
$\mathbb{R}^{1,3}$ with signature $(-+++)$.  
There are also a variety of nontrivial fluxes turned on in this
solution.  There is a five form flux
\be
F_5 = dC_4 + \star dC_4 \;,\, \mbox{where}
\; C_4 = \frac{1}{g_s h} dx^0 \wedge dx^1 \wedge dx^2 \wedge dx^3 \ ,
\ee
and where $g_s$ is the string coupling constant.
The finite smoothing indicates the presence of a harmonic $(2,1)$-form
$\omega_{2,1}$ which we take to be  imaginary 
self-dual: ${\star_X} \omega_{2,1} = i \omega_{2,1}$.
From $\omega_{2,1}$ we construct a three-form flux
$
G_3 = C \omega_{2,1}
$ where $C$ is a constant related to the rank of the gauge group in the dual theory.
The warp factor satisfies the relation
\be
\lap_X h = - \frac{g_s^2}{12} G_{abc} (G^*)^{\widetilde{abc}} \ ,
\ee
where the indices on $G_3$ are raised and the Laplacian $\lap_X$ is constructed
using the line element $ds_X^2$ without the warp factor. 
Although this solution holds for general $X$, clearly in order to know detailed behavior
of the fluxes and warp factor as a function of $p$, we need to know a metric on $X$.
  
From this perspective, we have chosen to study $dP_3$ and not $dP_2$ in this paper because $dP_3$ is known
to have a \K-Einstein metric \cite{TianYau} while $dP_2$ does not \cite{Matsushima}.
Given that $dP_3$ is \K-Einstein, we can construct a singular Calabi-Yau cone 
over $dP_3$ in a straightforward manner:
\be
ds_X^2 = dr^2 + r^2 \left[
(d\psi + \sigma)^2 + ds_V^2 \right] \ ,
\ee
where $\sigma = -2i (\partial \F - \bar \partial \F)$ and $\F$ is half the \K potential on $V = dP_3$.
Here $ds_V^2$ is
a \K-Einstein line element on $dP_3$.  In a hopefully obvious notation, $r$ is the radius
of the cone and $\psi$ an angle.
Although such a cone over $dP_2$ probably
exists as well, it will involve an irregular Sasaki-Einstein manifold as an intermediate
step; the metric on the Sasaki-Einstein manifold over $dP_2$ is not yet known.

Given that $dP_3$ is \K-Einstein,
the problem of finding $\omega_{2,1}$ on the 
singular cone reduces to finding a harmonic $(1,1)$-form $\theta$ on $dP_3$ such
that $\theta \wedge \omega = 0$ (where $\omega$ is the \K form on $V$) 
and ${\star_V} \theta = -\theta$, 
as pointed out in \cite{Franco:2004jz}.  
The relation between $\omega_{2,1}$ and $\theta$ is
 $\omega_{2,1} = (-i dr/r + d \psi + \sigma) \wedge \theta$.  

In addition to finding a numerical \K-Einstein metric on $dP_3$, we will also 
find a numerical $(1,1)$-form $\theta$, thus yielding a singular generalization of the 
KS solution for $dP_3$.  Historically, before the KS solution, Klebanov and Tseytlin \cite{KT}
derived exactly such a singular solution for the singular conifold.  
Although we have not produced a numerical solution for $h(p)$, 
with the explicit metric and numerical $(1,1)$-form for $dP_3$ in hand, we
have all the necessary ingredients to calculate $h(p)$.
The KT solution for $dP_1$ is known \cite{HEK}.

The next step would be to find a Ricci flat metric and imaginary
self-dual (2,1)-form on the smoothed cone over $dP_3$,
thus providing a generalization of the KS solution.  Such a solution would 
open up many future directions of study, both in gauge theory
and cosmology.  To name a handful of possibilities, one could compute
$k$-string tensions of the confining low energy gauge theory dual to
this $dP_3$ background, generalizing work of \cite{HK}.  
Alternatively, treating the SUGRA solution as a cosmology,
one could compute annihilation cross sections of 
cosmic strings \cite{Copeland:2003bj}
or slow roll parameters for D-brane inflation \cite{Kachru:2003sx}.

\section{Mathematical background}
\label{sec:mathbackground}

\subsection{Complex and symplectic coordinates}

The formalism we use to construct our numerical K\"ahler-Einstein metric
on $dP_3$ is based on work by Guillemin \cite{Guillemin} and later 
developed by Abreu \cite{Abreu}. Here we summarize this
formalism.

We consider an $n$ (complex) dimensional compact \K manifold $\Mm$. The manifold 
is equipped with the following three tensors:
\begin{itemize}
\item A complex structure ${J^\mu}_\nu$ satisfying ${J^\mu}_\nu{J^\nu}_\lambda = -{\delta^\mu}_\lambda$.
\item A symplectic form (also called in this context a \K form) $\omega$, which is a non-degenerate closed two-form.
\item A positive-definite metric $ds^2=g_{\mu\nu}dx^\mu dx^\nu$.
\end{itemize}
These tensors are related to each other by:
\begin{equation}
g_{\mu\nu} = \omega_{\mu\lambda}{J^\lambda}_\nu.
\end{equation}

Now let $\Mm$ also be a toric manifold. Tensors which are invariant under its $U(1)^n=T^n$ group of diffeomorphisms we call toric. In particular, we will restrict our
attention to toric metrics. Let $\Mm^\circ$ be the subset of $\Mm$ which is acted on freely by that group. There is a natural set of $n$ complex coordinates $z = u + i \theta$ on $\Mm^\circ$, where $u\in\R^n$ and $\theta\in T^n$ ($\theta_i\sim\theta_i+2\pi$); the $U(1)^n$ acts on $\theta$ and leaves $u$ fixed. Given that the manifold is K\"ahler, the metric may locally be expressed in terms of the Hessian of a K\"ahler potential $\F(z)$.\footnote{We follow the conventions of Abreu \cite{Abreu}. Note that $f$ is one-half the usual definition of the \K potential.}  Because $g_{\mu \nu}$ is invariant under the action of $U(1)^n$, the potential can be chosen to be a function of $u$.  The line element is
\be
ds^2 = g_{i \bar \jmath} \, dz_i d \bar z_{j} 
 + g_{\bar \imath j} d\bar z_i d z_j
= \Fij \, (du_i du_j + d\theta_i d\theta_j),
\ee
where we have introduced $F_{ij}(u)$:
\be
g_{i \bar \jmath} = 2\frac{\partial^2 \F}{\partial z_i \partial \bar z_j} =
\frac{1}{2} \frac{\partial^2 \F}{\partial u_i \partial u_j} = \frac{1}{2} \Fij \ .
\ee
The K\"ahler form is
\be
\omega = 2i\partial\bar\partial f=i g_{i \bar \jmath} \, dz_i \wedge d\bar z_j = 
\Fij \,  du_i \wedge d\theta_j\ .
\label{Kform}
\ee
The complex structure in these coordinates is trivial:
\be
-{J^{u_i}}_{\theta_j} = {J^{\theta_i}}_{u_j} = \delta^i_j, \qquad 
{J^u}_u={J^\theta}_\theta=0 \ .
\ee

It is often convenient to work with symplectic coordinates $w = x + i \theta$, which are related to the complex coordinates by:
\be\label{momentmap}
x \equiv \frac{\partial \F}{\partial u} \ .
\ee
Under this map (also known as the moment map), $\mathbb{R}^n$ is mapped to the interior $P^\circ$
of a convex polytope $P \subset \mathbb{R}^n$ which is given by the 
intersection of a set of linear inequalities, 
\be
P = \{x: l_a(x)\ge0 \ \forall a \} \ , \qquad l_a(x)=v_a \cdot x+\lambda_a \ ;
\ee
the index $a$ labels the faces, and the normal vector 
$v_a$ to each face is a primitive element of $\mathbb{Z}^n$.
These $v_a$ define the toric fan in the complex coordinates $z$.
We will see that the $\lambda_a$ determine the K\"ahler class of the metric. In terms of these new coordinates, the K\"ahler form becomes trivial:
\be
\omega = dx_i \wedge d \theta_i \ .
\label{Kformsymp}
\ee
Introducing the symplectic potential, which is the Legendre transform of $\F$,
\be
\G(x) = u\cdot x - \F(u) \ ,
\ee
the line element can be written
\be
ds^2 = \Gij \, dx_i dx_j + G^{ij} \, d\theta_i d\theta_j\ ,
\label{sympmetric}
\ee
where
\be
\Gij = \frac{\partial^2 \G}{\partial x_i \partial x_j},
\ee
and $G^{ij}$ is the inverse of $G_{ij}$. Note that, while $F_{ij}$ is regarded as a function of $u$ and $G^{ij}$ as a function of $x$, under the mapping \eqref{momentmap} the two matrices are equal to each other. The complex structure in symplectic coordinates is given by:
\be
{J^{x_i}}_{\theta_j} = -\Ginv, \qquad
{J^{\theta_i}}_{x_j} = \Gij, \qquad 
{J^x}_x={J^\theta}_\theta=0 \ .
\ee

To summarize, 
the complex and symplectic coordinate systems are related to each other by a Legendre transform:
\begin{equation}\label{legendre}
x = \pd{f}{u}, \qquad u = \pd{g}{x}, \qquad \F(u) + \G(x) = u \cdot x, \qquad \Fij(u) = \Ginv(x).
\end{equation}
Note that the whole system (complex and symplectic coordinate systems) has four gauge invariances under which the metric is invariant and the relations \eqref{legendre} are preserved:
\begin{enumerate}
\item $\F(u) \to \F(u)+c$, $\G(x) \to \G(x)-c$ for any constant $c$;
\item $\F(u) \to \F(u)+k \cdot u$, $x \to x+k$ for any vector $k$;
\item $\G(x) \to \G(x)+k \cdot x$, $u \to u+k$ for any vector $k$;
\item $x \to {M} \cdot x$, $u^t \to u^t \cdot M^{-1}$ for any element $M \in GL(n,\Z)$ (the ring $\mathbb{Z}$ is necessary to preserve the integrality of the boundary vectors $v_a$).
\end{enumerate}
However, gauge invariances (2) and (4) are broken by the polytope down to the subgroup of $\R^n\rtimes GL(n,\Z)$ under which $P$ is invariant.

\subsection{Boundary conditions and the canonical metric}
\label{sec:boundary}

The complement of $M^\circ$ in $M$ consists of points where one or more circles in the $T^n$ fiber degenerate. In order to have a smooth metric on all of $M$, there are two boundary conditions that must be imposed on $f(u)$, or equivalently $g(x)$. These are somewhat easier to express in the symplectic coordinate system. The first condition is that, as we approach a face of the polytope, the part of the metric parallel to the face, should not degenerate (or become infinite). Technically the requirement is that the function
\begin{equation}
\det G_{ij}\prod_al_a,
\end{equation}
which is positive in the interior of the polytope, should extend to a smooth positive function on the entire polytope. The second condition is that the shrinking circle(s) should go to zero size at the correct rate in order to avoid having a conical singularity. To express this boundary condition in terms of $g$, Guillemin \cite{Guillemin} and Abreu \cite{Abreu} introduced the \emph{canonical} symplectic potential:
\begin{equation}\label{canpotl}
\Gcan \equiv \frac12\sum_a l_a\ln l_a.
\end{equation}
This canonical potential leads to a metric on $M$ that is free of conical singularities. Furthermore, every smooth metric corresponds to a $g$ that differs from $\Gcan$ by a function that is smooth on the entire polytope; we will call this function $h$:
\begin{equation}\label{hdef}
\G = \Gcan + h.
\end{equation}

So far, we considered a toric manifold with a fixed metric. In general, a given toric manifold will admit many different toric metrics, i.e.\ \K metrics invariant under the given $U(1)^n$ diffeomorphism group. These will be described by functions $g(x)$ (or $f(u)$) that differ by more than the gauge transformations listed above. Metrics with the same polytope (modulo the gauge transformations 2 and 4, which act on the polytope) are in the same \K class. In general, the topology of the manifold is determined by the number and angles of the polytope's faces, i.e.\ the vectors $v_a$ (modulo gauge transformation 4), while the \K class is determined by their positions, i.e.\ the numbers $\lambda_a$ (modulo gauge transformation 2). If the $\lambda_a$ are all equal, then the \K class is proportional to the manifold's first Chern class. More specifically, if $\lambda_a=\Lambda^{-1}$ for all $a$, then $\Lambda[\omega] = 2\pi c_1(M)$.   The case of interest, $dP_3$, has $c_1(M)>0$, so we must take $\Lambda >0$.

\subsection{Examples}

In one complex dimension, there is only one compact toric manifold, $\mathbb{CP}^1$. The corresponding polytope $P$ is simply the interval, whose length determines the \K modulus. We take $P=[-\lambda,\lambda]$. The canonical symplectic potential
\begin{equation}
\Gcan(x) = \frac12\left((\lambda+x)\ln(\lambda+x)+(\lambda-x)\ln(\lambda-x)\right)
\end{equation}
yields the round metric of radius $\sqrt\lambda$:
\begin{equation}
ds^2_{\rm can} = \frac\lambda{\lambda^2-x^2}dx^2+\frac{\lambda^2-x^2}\lambda d\theta^2.
\end{equation}
The \K coordinate $u$ is related to $x$ by
\begin{equation}
x = \lambda\tanh u,
\end{equation}
and the \K potential is
\begin{equation}
f_{\rm can}(u) = \lambda\ln\cosh u,
\end{equation}
giving the metric in the form
\begin{equation}
ds^2_{\rm can} = \lambda\sech^2u(du^2+d\theta^2).
\end{equation}

\FIGURE{\epsfig{file=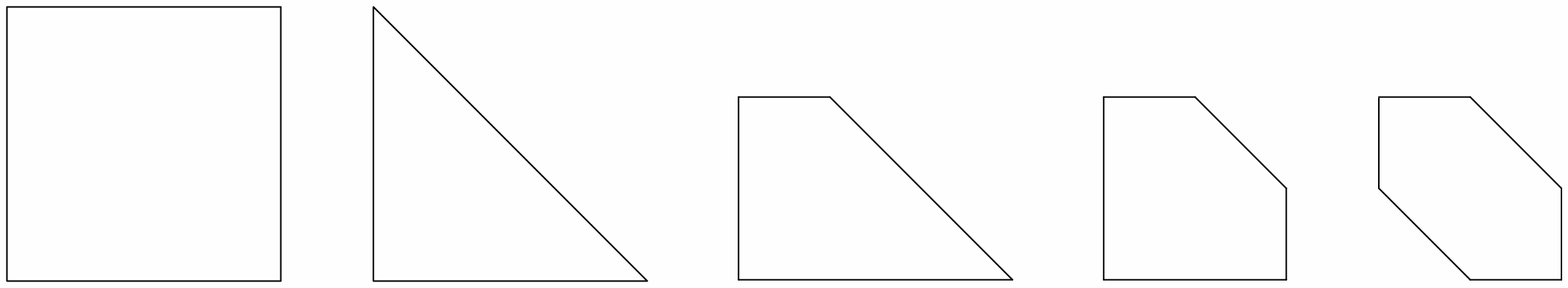,width=6in}\caption{The polytopes for the five compact toric manifolds with positive first Chern class: from left to right, $\CP^1\times\CP^1$, $\CP^2$, $dP_1$, $dP_2$, $dP_3$. Each polytope is drawn such that the $\lambda_a$ for all its faces are equal, corresponding to the \K class being proportional to the first Chern class.}\label{tfs}}

There are five compact toric surfaces with positive first Chern class (i.e.\ toric Fano surfaces); their polytopes are shown in Fig.~\ref{tfs}. $\mathbb{CP}^1\times 
\mathbb{CP}^1$ has two moduli, the sizes of the two $\mathbb{CP}^1$ factors; when these are equal the canonical metric is Einstein. $\mathbb{CP}^2$ has only a size modulus, and again the canonical metric is Einstein, as shown in Appendix \ref{sec:projective}. The del Pezzo surfaces $dP_1$, $dP_2$, and $dP_3$ have two, three, and four moduli respectively. Their canonical metrics are never Einstein. Indeed, $dP_1$ and $dP_2$ do not admit K\"ahler-Einstein metrics at all \cite{Matsushima}, essentially because their Lie algebra of holomorphic one-forms is not reductive. The case of $dP_3$ is special. On the one hand, it is known to admit a toric \K-Einstein metric \cite{TianYau}. On the other hand, that metric is not the canonical one; indeed it is not known in closed form --- hence the necessity of computing it numerically. $dP_3$ is discussed in more detail in Section \ref{sec:details} below.

\subsection{The Monge-Amp\`ere equation}
\label{sec:MA}

As usual for a K\"ahler manifold, the Ricci curvature tensor $R_{i\bar \jmath}$ 
can be written in complex coordinates $z$ in a simple way, namely
\be
R_{i \bar \jmath} = - \frac{\partial^2}{\partial z_i\partial\bar z_j} \ln \det g_{k \bar l} \ .
\label{Rdef}
\ee
Recall that there is a two-form ${\mathcal R} = i R_{i \bar \jmath} dz^i\wedge d \bar z^j$
associated with $R_{i \bar \jmath}$ where the class 
$[{\mathcal R}] = 2 \pi c_1 (M)$.
We are interested in K\"ahler-Einstein metrics on $\Mm$, that is metrics which
satisfy the following relation
\be
R_{i \bar \jmath} = \Lambda g_{i \bar \jmath}
\label{KErel}
\ee
for some fixed $\Lambda$. As explained above, this implies that $\lambda_a = \Lambda^{-1}$ for all $a$. The sign of $\Lambda$ is not arbitrary but is fixed by the first
Chern class of $\Mm$, which we are now assuming to be positive.

Given (\ref{Rdef}), we may integrate (\ref{KErel}) twice, yielding
\be
\ln\det \Fij = - 2 \Lambda \F + \gamma \cdot u - c
\label{KErel2}
\ee
where $\gamma$ and $c$ are integration constants.
In symplectic coordinates, this becomes
\be
\ln\det \Gij = -2 \Lambda \G +\frac{\partial \G}{\partial x} \cdot (2 \Lambda x - \gamma ) + c\ .
\label{GPDE}
\ee
Both \eqref{KErel2} and \eqref{GPDE} are examples of Monge-Amp\`ere type equations, which must be solved with the boundary conditions discussed in Section \ref{sec:boundary} above. The values of the constants $\gamma$ and $c$ in these equations are arbitrary; given a solution with one set of values, a solution with any other set can be obtained using gauge transformations (1) and (2). However, as discussed above, gauge transformation (2) acts on the polytope $P$ by a translation. Therefore, if we fix the position of $P$, there will be a unique $\gamma$ such that \eqref{GPDE} admits a solution.

Since the boundary conditions are non-standard, it is worth exploring them in more detail. Again, we work in the symplectic coordinate system. Recall that the condition for a smooth metric was that $h(x)$, defined by
\begin{equation}
g = \Gcan + h,
\end{equation}
be smooth on $P$, including on its boundary. We can re-write \eqref{GPDE} in terms of $h$ as follows:
\begin{equation}\label{hequation}
\ln\det\left(\delta_{ij} + G_{\rm can}^{ik}\frac{\partial^2h}{\partial x_k\partial x_j}\right) =
-2\Lambda h+\frac{\partial h}{\partial x} \cdot (2 \Lambda x - \gamma ) - \rho_{\rm can} + c\ ,
\end{equation}
where
\begin{equation}
\rho_{\rm can} \equiv
\ln\det G^{\rm can}_{ij} +2\Lambda g_{\rm can} - \pd{g_{\rm can}}{x}\cdot(2\Lambda x-\gamma).
\end{equation}
Normally, for a second-order PDE, we would expect to have to impose, for example, Dirichlet or Neumann boundary conditions in order to obtain a unique solution. However, in the case of \eqref{GPDE}, the coefficient of the normal second derivative goes to zero (linearly) on the boundary (near the face $a$ of the polytope, $G_{\rm can}^{ik}v_a^k\sim l_a$). Therefore, under the assumption that $h$ and its derivatives remain finite on the boundary, the equation itself imposes a certain (mixed Dirichlet-Neumann) boundary condition. If we were to try to impose an extra one, we would fail to find a solution. This is illustrated by the case of $\CP^1$. Setting $\Lambda = \lambda^{-1}$, we have $\rho_{\rm can}=0$, so that \eqref{hequation} becomes
\begin{equation}
\label{BCcpone}
\ln\left(1+\frac{\lambda^2-x^2}\lambda h''\right) = \frac2\lambda(-h+xh')+c;
\end{equation}
we see that the coefficient of $h''$ vanishes on the polytope boundary.

For future reference we record here the formulas for the Ricci and Riemann tensors in symplectic coordinates. Their non-zero components are
\begin{equation}\label{sympricci}
R_{x_ix_j} = R^{\theta_i\theta_j} =
\frac12\left(\frac{\partial^2}{\partial x_i\partial x_j} - G^{kl}\pd{G_{ij}}{x_k}\pd{}{x_l}\right)\ln\det G_{ij}
\end{equation}

\begin{equation}
R_{x_ix_jx_kx_l} = 
R_{x_ix_j}{}^{\theta_k\theta_l} = 
R^{\theta_i\theta_j}{}_{x_kx_l} = 
R^{\theta_i\theta_j\theta_k\theta_l} =
\frac12G^{mn}G_{ml[i}G_{j]nk}
\end{equation}
\begin{multline}
R{}^{\theta_i}{}_{x_j}{}^{\theta_k}{}_{x_l} =
-R{}_{x_i}{}^{\theta_j\theta_k}{}_{x_l} =
-R{}^{\theta_i}{}_{x_jx_k}{}^{\theta_l} =
R{}_{x_i}{}^{\theta_j}{}_{x_k}{}^{\theta_l} \\ =
\frac12\left(G_{ijkl}-G^{mn}G_{ijm}G_{kln}-G^{mn}G_{ml(i}G_{j)nk}\right),
\end{multline}
where we've defined
\begin{equation}
G_{ijk} \equiv \pd{G_{jk}}{x_i}, \qquad G_{ijkl} \equiv \pd{G_{jkl}}{x_i}.
\end{equation}

\subsection{Volumes}

Here follows a short discussion about volumes useful for understanding the relation between
$\lambda_a$ and $\Lambda$ above.

We know that the volume of a complex surface $M$  
is
\be
\Vol(\Mm) = \frac{1}{2} \int_{\Mm} \omega^2 \ ,
\label{volS}
\ee
while the volume of a curve $C$ is
\be
\Vol(C) = \int_C \omega \ .
\label{volC}
\ee
From the K\"ahler-Einstein condition (\ref{KErel}), (\ref{Kform}), and the fact
that the class of the Ricci form is related to the first Chern class,
$[\mathcal{R}] = 2 \pi c_1(\Mm)$, it follows that $[\omega] = 2\pi c_1(\Mm)/ \Lambda$ and
that
\be
\Vol(\Mm) = \frac{2\pi^2}{\Lambda^2} c_1(M)^2 \ .
\ee
For $\mathbb{CP}^2$ blown up at $k$ points, $c_1^2 = 9-k$. 
Meanwhile, for our curve,
\be
\Vol(C) = \frac{2\pi}{\Lambda} c_1(M) \cdot C \ .
\ee

In symplectic coordinates, it is easy to compute these volumes.  From
(\ref{Kformsymp}) or \eqref{sympmetric}, the volume (\ref{volS}) reduces to $4\pi^2$ times the
area of $P$.  Setting $\Lambda = 1$ corresponds to setting
the $\lambda_a=1$.
For curves, the computation is similarly easy.
For example, some simple torus invariant curves correspond to edges of $P$,
and the volume computation reduces to measuring the length of an edge of $P$.

\subsection{The Laplacian}

The Laplacian acting on a scalar function $\psi$ is,
\begin{equation}
\lap \psi = \frac{1}{\sqrt{\det{g_{\alpha\beta}}}} \partial_{\mu} \left[ \sqrt{\det{g_{\alpha\beta}}} \; g^{\mu\nu} \partial_{\nu} \psi \right] \ .
\end{equation}
In symplectic coordinates, the determinant of the metric $\sqrt{\det g_{\alpha \beta}} = 1$.  We consider
a function $\psi$ invariant under the torus action, implying no dependence on the two angular coordinates so that $\psi = \psi(x_1,x_2)$.   Thus, the Laplacian can be written in a simpler fashion:
\be
\lap \psi = \frac{\partial}{\partial x_i}\left( \Ginv  \frac{\partial \psi}{\partial x_j} \right) \ .
\ee
The Laplacian thus depends on a third derivative of $\G$.
 However since we are interested in Einstein metrics, 
 we may take \eqref{GPDE}, differentiate it, and use it to eliminate this derivative. 
 One then finds the form
\begin{equation}
\lap_{\mathrm{E}} \psi = \Ginv_{\rm E} \frac{\partial^2 \psi}{\partial x_i \partial x_j} - 
2 \Lambda x_i \frac{\partial \psi}{\partial x_i} \ .
\label{einstlap}
\end{equation}
From this equation one deduces the interesting fact that the symplectic coordinates are eigenfunctions
of $-\lap_{\mathrm{E}}$ with eigenvalue $2 \Lambda$ --- the symplectic coordinates are close
to being harmonic coordinates.

\subsection{Harmonic $(1,1)$-forms}
\label{harmonicforms}

A compact toric variety $M$ of dimension $n$
defined by a fan $\{ v_a \}$ of $m$ rays has Betti number $b_{2} = m-n$
\cite{Fulton}.  Moreover, $dP_3$ has no holomorphic (nor antiholomorphic) 2-forms
and thus must have $m-n$ harmonic (1,1)-forms.
These (1,1)-forms are in one-to-one correspondence with torus invariant Weil divisors $D_a$ 
modulo linear equivalence. The equivalence relations are
\be
\sum_a v_a^i D_a  \sim 0 \ .
\ee

From their connection with the divisors $D_a$, perhaps it is not surprising that 
these (1,1)-forms $\theta_a$ can be constructed from functions $\mu_a$ which have a
singularity of the form $\ln(v_a \cdot x + 1)$ along the boundaries of the
polytope \cite{Abreu}.  Moreover, they should satisfy Maxwell's equations,
\be
d\theta = 0 \; {\rm{and}} \; d\star \theta =0 \ .
\ee
The first equation $d\theta=0$ is immediate from the local description of
$\theta_{i \bar \jmath}$ as
$\partial_i \partial_{\bar \jmath} \mu$.  The other equation is
\be
0=D^i \theta_{i \bar \jmath} = g^{\bar k i} D_{\bar k} \theta_{i \bar \jmath} \ .
\ee
Using the Bianchi identity, we find then
\be
0 = g^{\bar k i} D_{\bar \jmath} \theta_{i \bar k}
= \partial_{\bar \jmath}\left( g^{i \bar k} \theta_{i \bar k} \right) \ ,
\ee
or
\be
g^{i \bar k} \theta_{i \bar k} = \rm{constant} \ .
\ee
Now this last equation is rather interesting.  The left  hand side
is the Laplacian operator acting on $\mu$.  In symplectic coordinates,
the left hand side can be rewritten to yield
\be
\frac{\partial}{\partial x_i} \left( \Ginv \frac{\partial \mu }{\partial x_j}  \right) =
\mbox{constant} \ .
\ee
Using the K\"ahler-Einstein condition, the left hand side becomes
\eqref{einstlap}.

The constant is easy to establish.  We have that
\be
\Vol(M) g^{i \bar \jmath} \theta_{i \bar \jmath} = \int_M \theta \wedge \star \omega \ .
\ee
The \K form is self-dual under the Hodge star, $\star \omega = \omega$.
Conventionally, we may write $[\theta] = 2\pi c_1(D)$, assuming $\theta$
is the curvature of a line bundle ${\mathcal O}(D)$.  Using finally that
$[\omega] = 2\pi c_1(M)$ (for $\Lambda=1$), we find that
\be
g^{i \bar \jmath} \theta_{i \bar \jmath} = 2 \frac{D\cdot K}{K^2} \ ,
\ee
where $K$ is the canonical class of $M$.  This constant is often referred to as the slope
of ${\mathcal O}(D)$.

\subsection{More on $dP_3$}
\label{sec:details}

For $dP_3$, the toric fan is described by the six rays spanned by 
\be
v_1 = (1,0) \; ; \; \; \;
v_2 = (1,1) \; ; \; \; \;
v_3 = (0,1) \; ; 
\ee
\[
v_4 = (-1,0) \; ; \; \; \;
v_5 = (-1,-1) \; ; \; \; \;
v_6 = (0,-1) \ . 
\]
This fan leads to a dual polytope $P$ which is a hexagon. As discussed above, when all six $\lambda_a$ are equal, the \K class is proportional to the first Chern class; we will choose $\lambda_a=1$. In this case the polytope has a dihedral symmetry
group $D_6$, generated by the following $\mathbb{Z}_2$ reflection and $\mathbb{Z}_6$ rotation:
\be
R_1 = \begin{bmatrix} 0 & 1 \\ 1 & 0 \end{bmatrix}, \qquad R_2 = \begin{bmatrix} 1 & 1 \\ -1 & 0 \end{bmatrix} \ .
\label{dihedral}
\ee
This discrete symmetry will be shared by the \K-Einstein metric on $dP_3$, and will therefore play an important role in our computations. Note that the element $R_2^3$ acts as $x\to-x$, which sets $\gamma=0$ in \eqref{KErel2} and \eqref{GPDE}. Note also that the $D_6$ acts naturally on the group of Cartier divisors
on $dP_3$ (and hence on the harmonic (1,1)-forms), as
can easily be seen by thinking of the set of $v_a$ as divisors on the manifold.

The hexagon $P$ has a natural interpretation as the intersection of the polytope for a symmetric unit $(\mathbb{CP}^1)^3$, which is a cube with coordinates $x_1,x_2,x_3$ satisfying $|x_i|\le1$, with the plane $x_1+x_2+x_3=0$. The intersection defines a natural embedding of $dP_3$ into $(\mathbb{CP}^1)^3$. Furthermore, the canonical metric on $dP_3$ is simply the one induced from the Fubini-Study metric on $(\mathbb{CP}^1)^3$.

While the action of the symmetries in symplectic coordinates is geometrically clear, we want to make explicit their description in complex coordinates.
Note that the action of $R_2$ on the complex coordinates $u_i$ is given by the inverse transpose of $R_2$ since $u=\partial g/\partial x$. 
Thus $R_2$ acts as $(u_1,u_2) \mapsto (u_2,u_2-u_1)$. Similarly, $R_1$ which is its own inverse transpose, acts by $(u_1,u_2) \mapsto (u_2,u_1)$. 
More explicitly, the relation between the complex and symplectic coordinates for the
canonical potentials is given by 
\be
X^2 = e^{2u_1} = \frac{1+x_1}{1-x_1} \frac{1+x_1+x_2}{1-x_1-x_2} \ ,
\ee
\[
Y^2 = e^{2u_2} = \frac{1+x_2}{1-x_2} \frac{1+x_1+x_2}{1-x_1-x_2} \ .
\]
To invert these relations involves solving a cubic equation in one of the $x_i$.

There are six affine coordinate patches associated to the cones formed by pairs
of neighboring rays $v_a$.  We denote the cone formed by
the ray $v_a$ and $v_{a+1}$ to be $\sigma_a$ and the coordinate system on $\sigma_a$ by
$(\xi_a, \eta_a)$.  The complex coordinates on the patches $a=1, \ldots , 6$
are 
\be
\begin{array}{cl}
\sigma_1: & (\xi_1, \eta_1) = (Y, X Y^{-1}) \\
\sigma_2: & (\xi_2, \eta_2) = (X^{-1} Y, X) \\
\sigma_3: & (\xi_3, \eta_3) = (X^{-1}, Y) \\
\sigma_4: & (\xi_4, \eta_4) = (Y^{-1}, X^{-1} Y) \\
\sigma_5: & (\xi_5, \eta_5) = (XY^{-1}, X^{-1}) \\
\sigma_6: & (\xi_6, \eta_6) = (X, Y^{-1}) 
\end{array}
\label{xietachart}
\ee
Note that $\eta_a = 1/\xi_{a+1}$, and that when $\xi_a = \eta_{a+1}$, then
$\xi_{a+1} = \eta_a$.  Because of these relations, we can consider an atlas
on $dP_3$ where for each $\sigma_a$, we restrict $\xi_a \leq 1$ and $\eta_a \leq 1$.
These six polydisks $P_a$ tile $dP_3$.  

In the symplectic coordinate system $(x_1, x_2)$, our atlas
divides the hexagon up into six pieces.  The boundaries are given by
the conditions $e^{u_1} = 1$, $e^{u_2} = 1$, and $e^{u_1-u_2} = 1$ or in
symplectic coordinates, the three lines $2x_1 + x_2=0$, $x_1 + 2x_2=0$, 
and $x_1 -x_2=0$.

The action of $D_6$ maps one $P_a$ into another.  For example in patch 
$3$ since $X$ and $Y$ are the exponentials 
of $u_1$ and $u_2$, $R_2$ acts on our coordinates  
by sending $(X^{-1},Y)\to (Y^{-1}, X^{-1} Y)$, mapping 
patch $3$ into patch $4$.  

\FIGURE{
\centerline{\psfig{figure=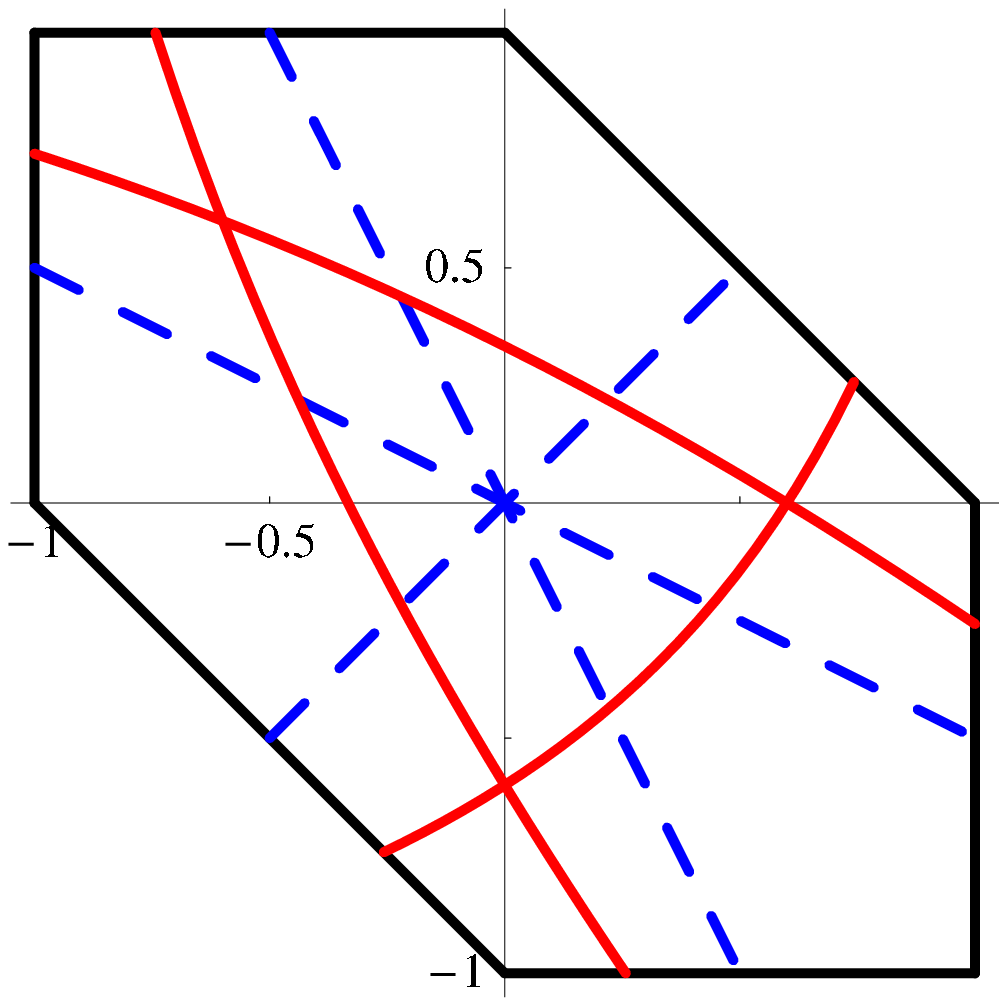, width=2.5in}}
\caption{ The polytope $P$ for $dP_3$.  The dashed lines correspond to the
edges of the unit polydisks: $2x_1 + x_2 = 0$, $x_1+2x_2=0$, and $x_1=x_2$.
The red curved lines correspond to setting $\xi_i$ and $\eta_i$ equal to two instead
of one in the appropriate coordinate systems.}
\label{patches}
}

\section{Ricci flow}
\label{sec:ricciflow}

In this section we consider the flow, on the space of \K metrics on $M$, defined by \eqref{RFdef1}:
\begin{equation}\label{RFdef}
\pd{g_{\mu\nu}}{t}=-2R_{\mu\nu}+2g_{\mu\nu},
\end{equation}
where we have set $\Lambda=1$. Note that, since the Ricci form is a closed two-form, a \K metric remains \K along the flow. Furthermore, if $[\omega]=[{\mathcal R}]$ for the initial metric, then the flow will stay within that \K class. If the flow converges, the limiting metric will clearly be Einstein. A recent result of Tian and Zhu \cite{Tian} implies that, on $dP_3$, starting from any initial \K metric obeying $[\omega]=[{\mathcal R}]$, the flow will indeed converge to the \K-Einstein metric. 
Numerical simulation of the flow can therefore be used as an algorithm for finding the \K-Einstein metric. 

Thanks to the toric symmetry, \eqref{RFdef} can be written as a parabolic partial differential equation for a single function in 2 space and 1 time dimensions, and therefore represents a simple problem in numerical analysis. Specifically, according to \eqref{Rdef}, we can write \eqref{RFdef} in terms of the \K potential $f(u)$:
\begin{equation} \label{feqn}
\pd{\F}{t} =  \ln \det \Fij+2\F + c.
\end{equation}
Here $c$, which is independent of $u$ but may depend on $t$, can be chosen arbitrarily. In particular, the zero mode of $f$ (which is pure gauge) is clearly unstable according to \eqref{feqn}, and $c$ is useful for controlling that instability. In principle we could also add a term $\gamma\cdot u$ to the right-hand side. As explained in Section \ref{sec:details}, however, due to the symmetry of the polytope for $dP_3$, we know that $\gamma=0$ in the solution to the Monge-Amp\`ere Eq.~\eqref{KErel2}.
In the symplectic coordinates, \eqref{feqn} takes the form
\begin{equation} \label{geqn}
\pd{\G}{t} = \ln \det \Gij+2\left(-x \cdot \pd{\G}{x}+\G\right) - c.
\end{equation}
Note that, because the \K potential is $t$-dependent, the mapping relating $u$ to $x$ is as well. To derive \eqref{geqn}, we used the fact that, since $f$ and $g$ are related by a Legendre transform, $\partial g/\partial t|_x = -\partial f/\partial t|_u$. Note also that \eqref{geqn} does not quite describe Ricci flow in the symplectic coordinate system; rather it describes Ricci flow supplemented with a $t$-dependent diffeomorphism (described by the flow equation $\partial g_{\mu\nu}/\partial t=-2R_{\mu\nu}+2g_{\mu\nu}+2\nabla_{(\mu}\xi_{\nu)}$). Under pure Ricci flow, symplectic coordinates do not stay symplectic, since the Ricci tensor \eqref{sympricci} is not the Hessian of a function, hence the necessity of supplementing the flow with a diffeomorphism. 

The two flows \eqref{feqn} and \eqref{geqn} were simulated by two independent computer programs, which yielded consistent results. In both cases we represented the (\K or symplectic) potential using standard real space finite differencing, and therefore obtain the resulting approximation to the geometry in the form of the potential at an array of points. (In the next section we discuss how to present this information more compactly using polynomial approximations.) We relegate technical details of these implementations to Appendix \ref{sec:numerics}. When using finite difference methods it is important to show that one obtains a suitable convergence to a continuum limit upon refinement of the discrete equations, and we present tests that confirm this, and estimate errors in the same appendix.

\subsection{Complex coordinates}
\label{sec:complex}

We now discuss in more detail the implementation of Ricci flow using complex coordinates.
Our canonical K\"ahler potential $\Fcan$ is defined in $P^\circ$, the interior of the hexagon.
To work on our atlas (\ref{xietachart}), we use \K transformations to modify the K\"ahler potential so that
$\F$ is smooth on each coordinate patch, including the two relevant edges
of the hexagon.  
This K\"ahler transformation must also respect the 
$U(1)$ isometries so must only depend on the $u_i$. The only such K\"ahler 
transformations are affine linear combinations of the $u_i$. 
(For a function
$h(u_1+i \theta_1,u_2+i \theta_2)$ 
depending only on the $u_i$, 
the condition on \K transformations $\partial\bar{\partial} h \equiv 0$ 
reduces to $\frac{\partial^2 h}{\partial u_i\partial u_j}\equiv 0$.) 
In the interior of the hexagon, 
we can thus modify $\F$ without modifying the metric by adding a linear combination
of $u_1$ and $u_2$.  In symplectic coordinates
\begin{eqnarray*}
u \cdot \kappa = \frac{1}{2} \sum_{a=1}^6 v_a \cdot \kappa \ln (v_a \cdot x + 1)
\end{eqnarray*}
where $\kappa \in \mathbb{R}^2$.

We find that for any \K potential $f$, regardless of coordinate patch, 
\be
\det \Fij = \xi^2 \eta^2 \left[ \left(
		\frac{1}{\xi} \frac{\partial \F}{\partial \xi} +
		\frac{\partial^2 \F}{\partial \xi^2}
\right) 
\left( 
	\frac{1}{\eta} \frac{\partial \F}{\partial \eta} +
		\frac{\partial^2 \F}{\partial \eta^2}
\right) - \left( 
		\frac{\partial^2 \F}{\partial \xi \partial \eta}
\right)^2 \right] \ .
\ee
We can rewrite (\ref{feqn}), yielding
\be
\pd{f}{t} = \ln \left[ \left(
		\frac{1}{\xi} \frac{\partial \F}{\partial \xi} +
		\frac{\partial^2 \F}{\partial \xi^2}
\right) 
\left( 
	\frac{1}{\eta} \frac{\partial \F}{\partial \eta} +
		\frac{\partial^2 \F}{\partial \eta^2}
\right) - \left( 
		\frac{\partial^2 \F}{\partial \xi \partial \eta}
\right)^2 \right]   + 2 (f + \ln \xi + \ln \eta) +c  \ .
\label{KErelmod}
\ee
This shift of $f$ by logarithms is a \K transformation.  
In each coordinate patch, we can write
$\ln \xi_a + \ln \eta_a = u \cdot \kappa_a$ where 
$\kappa_1 = (1,0)$, $\kappa_2 = (0,1)$, $\kappa_3 = (-1,1)$, 
$\kappa_4 = (-1,0)$, $\kappa_5 = (0,-1)$, and $\kappa_6=(1,-1)$.
Moreover, for these choices of $\kappa_a$, $f_a = \F + u \cdot \kappa_a$
is well behaved everywhere inside $P_a$ including the edges.

We simulate Ricci flow using a \K-Einstein potential $\F_a$ defined on a neighborhood $U_a$ of $P_a$,
$U_a = \{ (\xi_a, \eta_a) \; : \; \xi_a, \eta_a < L, L>1 \}$ which satisfies (\ref{KErelmod}).
As an initial condition, we take $f_a(t=0) = \Fcan + \ln \xi_a + \ln \eta_a$.
Given two patches, $a$ and $b$, then for a point in the overlap, 
$p \in U_a \cap U_b$,
\be
\F_b(p) = \F_a(p) + u \cdot (\kappa_{b} - \kappa_a) \ .
\ee
These \K transformations are consistent with the canonical potential and thus fix the same
\K class.
There is then an element $R \in D_6$ such that $R(P_b) = P_a$ which relates
$\F_a$ and $\F_b$. In particular, if $p$ in is in $U_b$ and $Rp$ is in $U_a$:
\be
\F_b(p) = \F_a(R p) \ .
\ee
Thus, with these quasiperiodic boundary conditions along the interior edges
$\xi_a = L$ and $\eta_a=L$, 
we can determine $\F$ by working solely on the patch $U_a$.  

In addition to quasiperiodic boundary conditions along the interior edges of $U_a$, 
close to the exterior boundary of $U_a$, we use the condition that
the normal derivative to the boundary must vanish.  These Neumann boundary conditions
arise because in complex coordinates, approaching the exterior boundary should be like
approaching the center of the complex plane. 

The Ricci flow in this domain was represented using second order accurate finite differencing with various resolutions up to a grid ${\mathcal F}_{IJ}$
of $250 \times 250$ points. This discretization, its convergence to the continuum and error estimates at this resolution are discussed in the Appendix \ref{sec:numerics}, and data presented in this section is given either from extrapolating to the continuum or using this highest resolution. In particular we estimate that the K\"ahler potential computed at this resolution is accurate pointwise to one part in $10^5$.

We noted earlier that while Ricci flow may converge, the zero mode of the potential $f$ is not guaranteed to converge since it does not appear in the metric. For convenience we wish to obtain a flow of $f$ that does converge, so we promote the constant $c$ to be time dependent along the flow, but still a constant on the geometry (i.e.~we introduce a flow dependent constant K\"ahler transformation), and choose
$c$ such that the value of $\Fd_{N/2, N/2}$, where $N$ is the grid size, does not change.  
At large times, $c$ tends to a constant. 
At the end of the flow,
we found it convenient to set $c=0$ by adding an appropriate constant value to the grid
$\Fd$.

\subsection{Symplectic coordinates}
\label{sec:symplectic}

We move on to discuss the implementation of Ricci flow based on symplectic coordinates. As discussed at the beginning of the section, although we work in symplectic coordinates, we use Ricci flow defined by the complex coordinates; since the relation between the two coordinate systems moves around as the metric changes, in the symplectic coordinates this is Ricci flow plus diffeomorphism. 

In order to deal with the boundary conditions on the edge of the polytope, it is useful to work with $h$ rather than $\G$ (recall that $\G=\Gcan+h$ and $\Gcan$ is the canonical symplectic potential). In terms of $h$, \eqref{geqn} becomes
\begin{equation}\label{heom}
\pd{h}{t} = 
\ln\det\left(\delta_{ij}+ G_\can^{ik}\frac{\partial^2h}{\partial x_k\partial x_j}\right)
+2\left(-x\cdot\pd{h}{x}+h\right)
+ \rho_\can - c,
\end{equation}
where
\begin{equation}
\rho_\can \equiv \ln\det G^{\rm can}_{ij}+2\left(-x \cdot \pd{\Gcan}{x}+\Gcan\right)
= \ln(L_1+L_2+L_3),
\end{equation}
\begin{equation}
G_\can^{ik} = 
\frac{L_1L_2}{L_1+L_2+L_3}\begin{bmatrix}1+\frac{L_3}{L_2} & -1 \\ -1 & 1+\frac{L_3}{L_1}\end{bmatrix},
\end{equation}
\begin{equation}
L_1 = l_1l_4 = 1-x_1^2, \qquad L_2 = l_3l_6 = 1-x_2^2, \qquad L_3 = l_2l_5 = 1-(x_1+x_2)^2.
\end{equation}

Due to the hexagon's $D_6$ symmetry, it is sufficient to simulate the flow within the square domain $0\le x_1\le1$, $-1\le x_2\le0$. Ricci flow in this domain was represented using second order accurate finite differencing with various resolutions up to a grid of $400\times400$ points. The differencing, continuum convergence and errors at this resolution are discussed in the Appendix \ref{sec:numerics} with data presented in this section given either from extrapolating to the continuum or using this highest resolution. In that appendix we estimate that the symplectic potential computed at this resolution is accurate to one part in $10^6$ at a given point.

As for the complex coordinates, the zero mode of the symplectic potential is pure gauge and does not converge. The constant $c$ was promoted to depend on flow time and chosen to keep $h(0,0,t)=0$.

\subsection{Results}
\label{sec:rfresults}

\FIGURE{
\epsfig{file=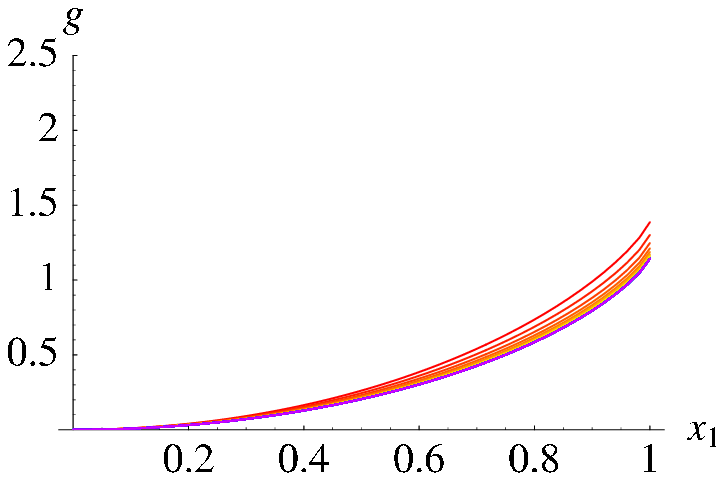,width=3in}\ \epsfig{file=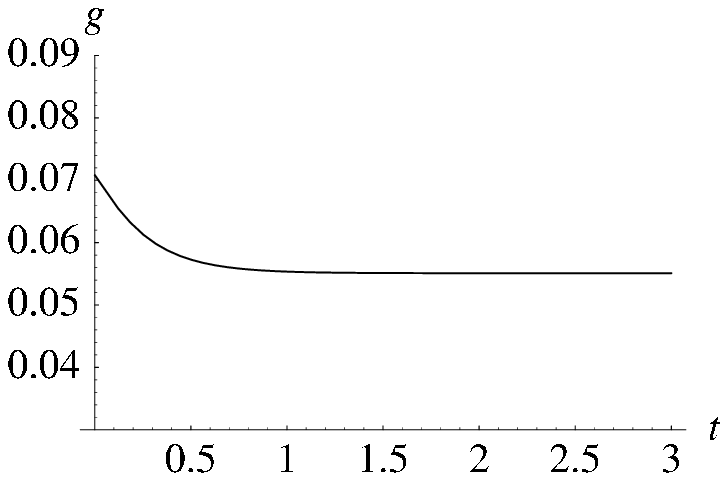,width=3in} \\
\epsfig{file=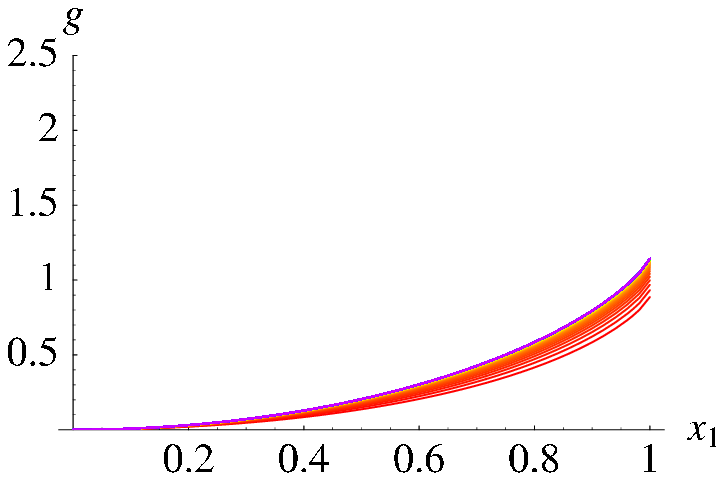,width=3in}\ \epsfig{file=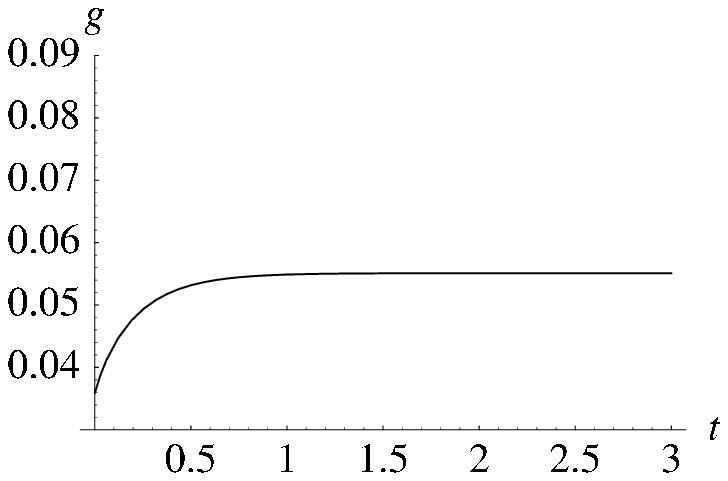,width=3in} \\
\epsfig{file=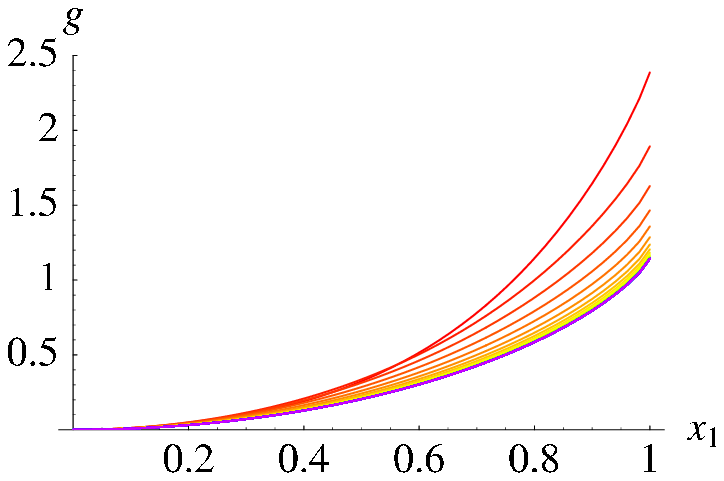,width=3in}\ \epsfig{file=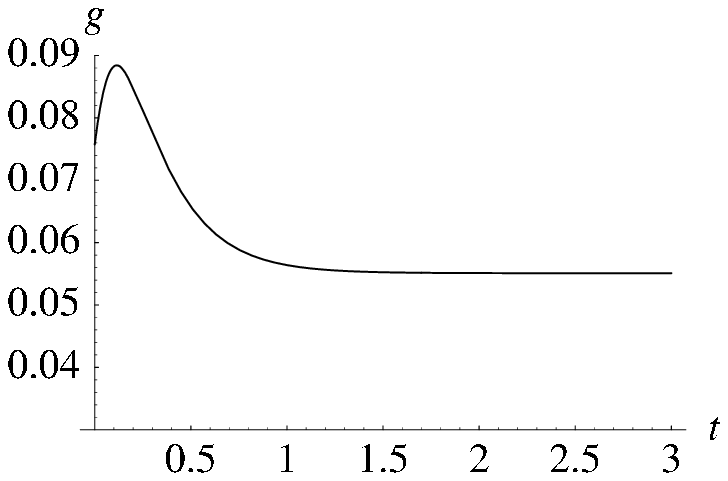,width=3in}
%\caption{The Ricci flow using symplectic coordinates starting with (left) the canonical metric ($h_0=0$), (middle) $h_0=-\frac12(x_1^2+x_2^2+x_1x_2)$, (right) $h_0=(x_1^2+x_2^2+x_1x_2)^2$. Top: $\G=\Gcan+h$ versus $x_1$ along the line $x_2=0$, plotted at intervals of 0.1 units of Ricci flow time ($t=0$ is the top curve). Bottom: $g$ versus $t$ at the (arbitarily chosen) point $(x_1,x_2)=(0.3,-0.2)$. We see that all three initial conditions converge to the same fixed point.}
\caption{The Ricci flow using symplectic coordinates starting with (top) the 
canonical metric ($h_0=0$), (middle) $h_0=-\frac12(x_1^2+x_2^2+x_1x_2)$, 
(bottom) $h_0=(x_1^2+x_2^2+x_1x_2)^2$. Left side: $\G=\Gcan+h$ versus $x_1$ 
along the line $x_2=0$, plotted at intervals of 0.1 units of Ricci flow time 
($t=0$ is the top curve). Right side: $g$ versus $t$ at the (arbitarily chosen) 
point $(x_1,x_2)=(0.3,-0.2)$. We see that all three initial conditions 
converge to the same fixed point.}
\label{tevol}
}

As predicted by the Tian-Zhu theorem, the metric converges smoothly and uneventfully to the K\"ahler-Einstein one. In the symplectic implementation, the Ricci flow was simulated starting with a variety of initial functions $h_0(x)=h(x,t=0)$ (always corresponding, of course, to positive-definite initial metrics). Three examples are shown in Fig.~\ref{tevol}. For every initial function investigated, the flow converged to the same fixed point $h_{\rm E}(x)=h(x,t=\infty)$, which necessarily represents the \K-Einstein metric. The exponential approach to the fixed point is controlled by the scalar Laplacian; this will be discussed in detail in the next subsection.

\FIGURE{
\centerline{a) \epsfig{file=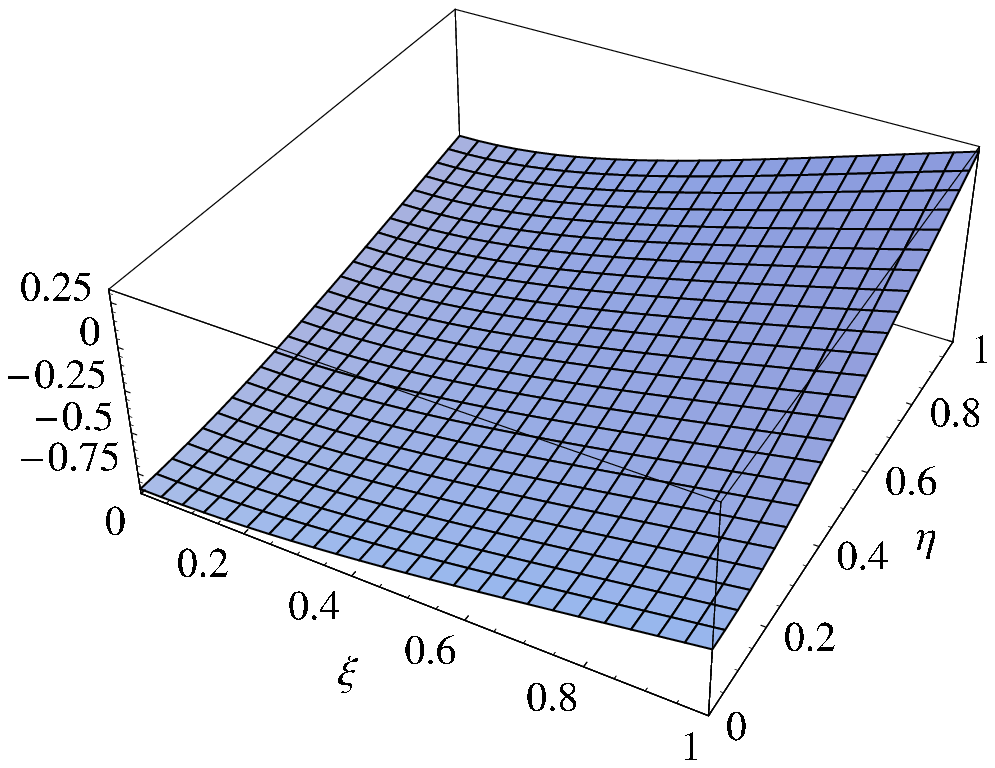, width=2in} \hskip 0.1in b)
\epsfig{file=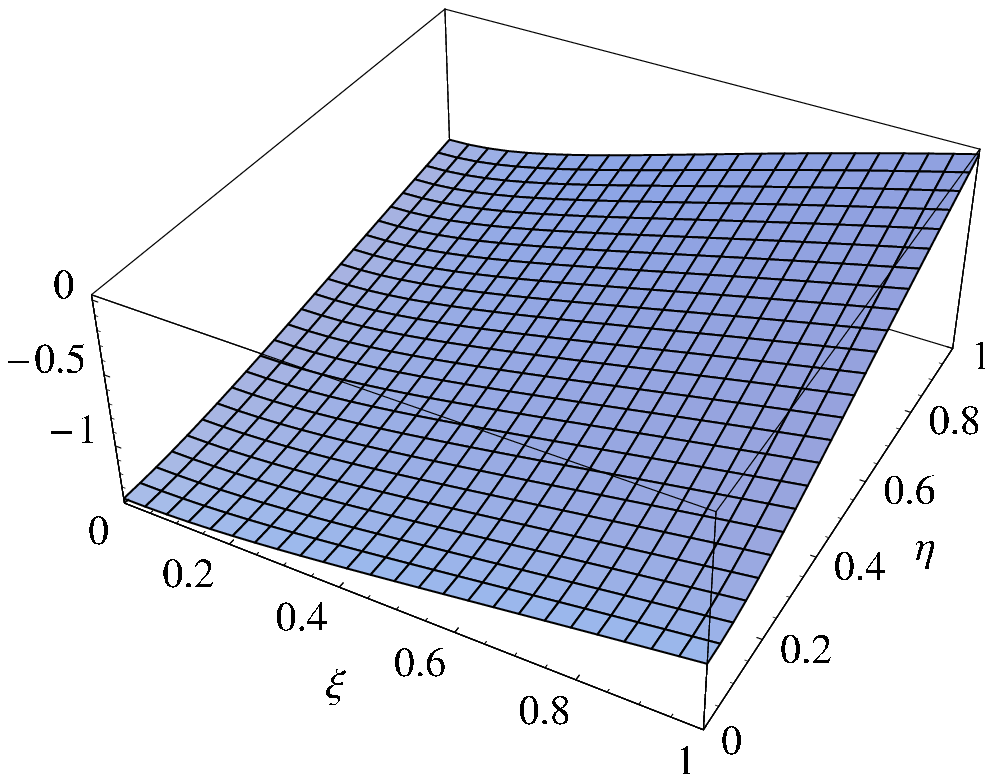, width=2in} \hskip 0.1in c) \epsfig{file=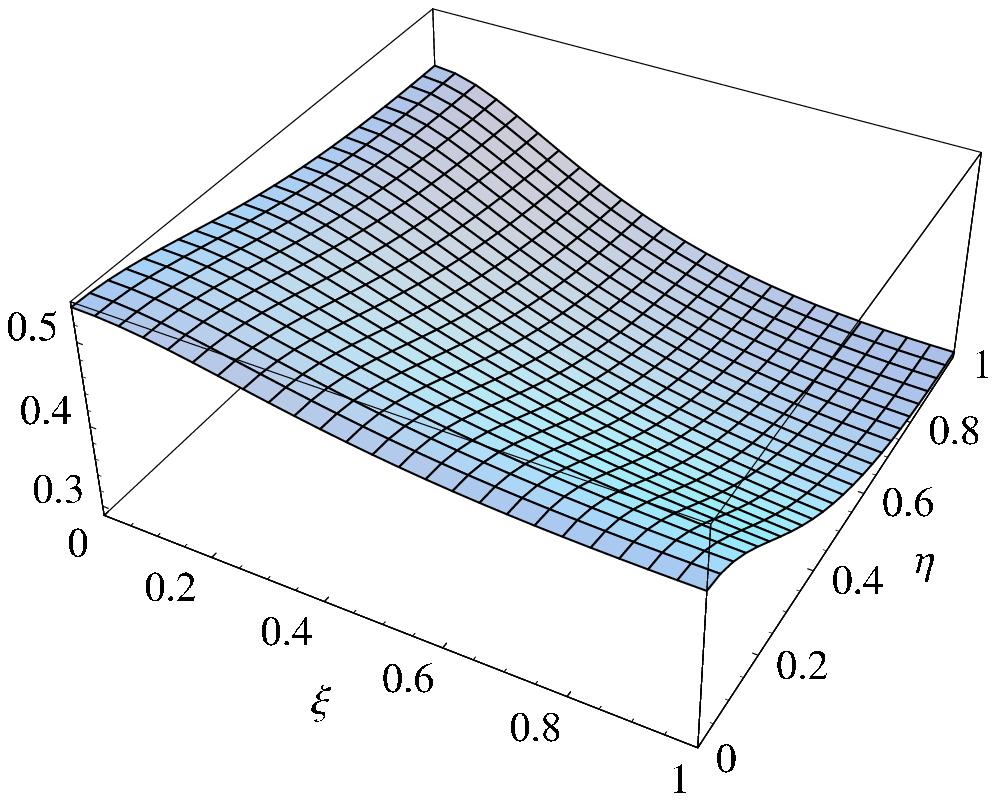, width=2in}}\\
\caption{a) The Einstein \K potential $\F_{\rm E}$ as a function of $\xi$ and $\eta$;
b) the canonical potential $\Fcan(\xi, \eta)$;
and c) $\F_{\rm E}(\xi, \eta) - \Fcan(\xi, \eta)$.}
\label{complexresults}
}
\FIGURE{
\epsfig{file=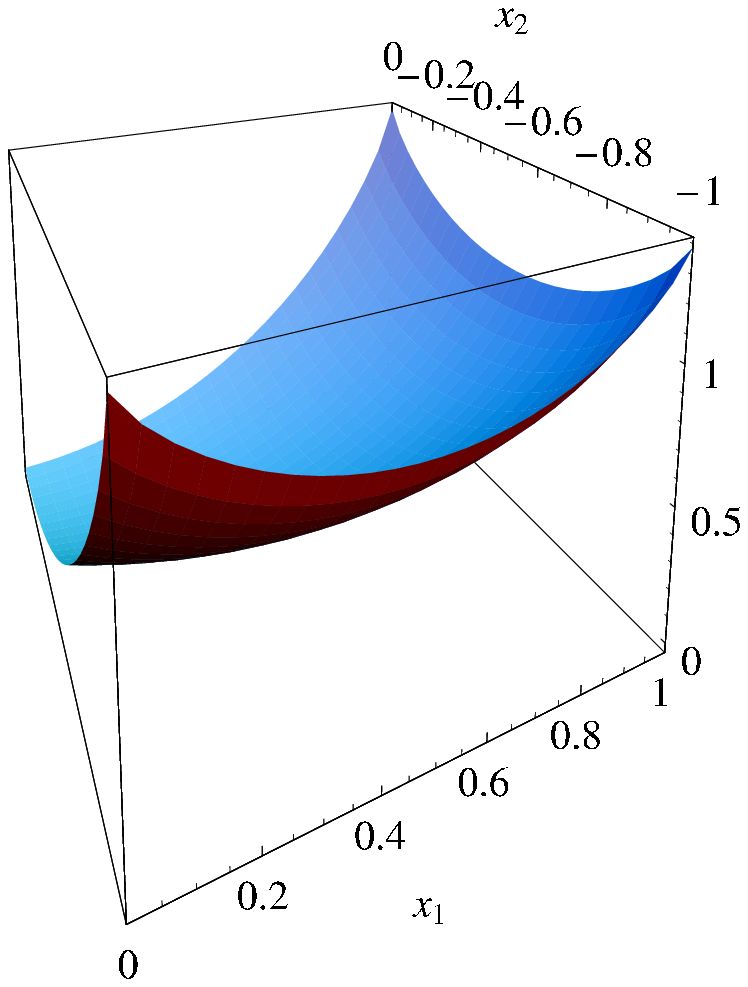,width=2.6in}\quad\epsfig{file=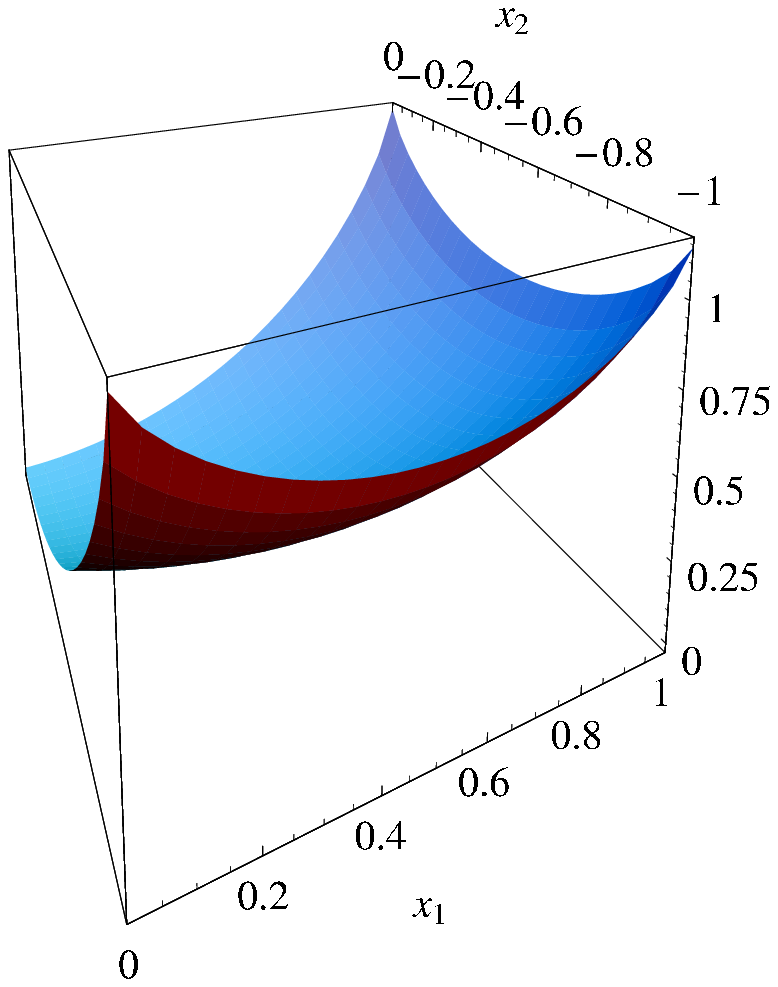,width=2.6in} \\
\epsfig{file=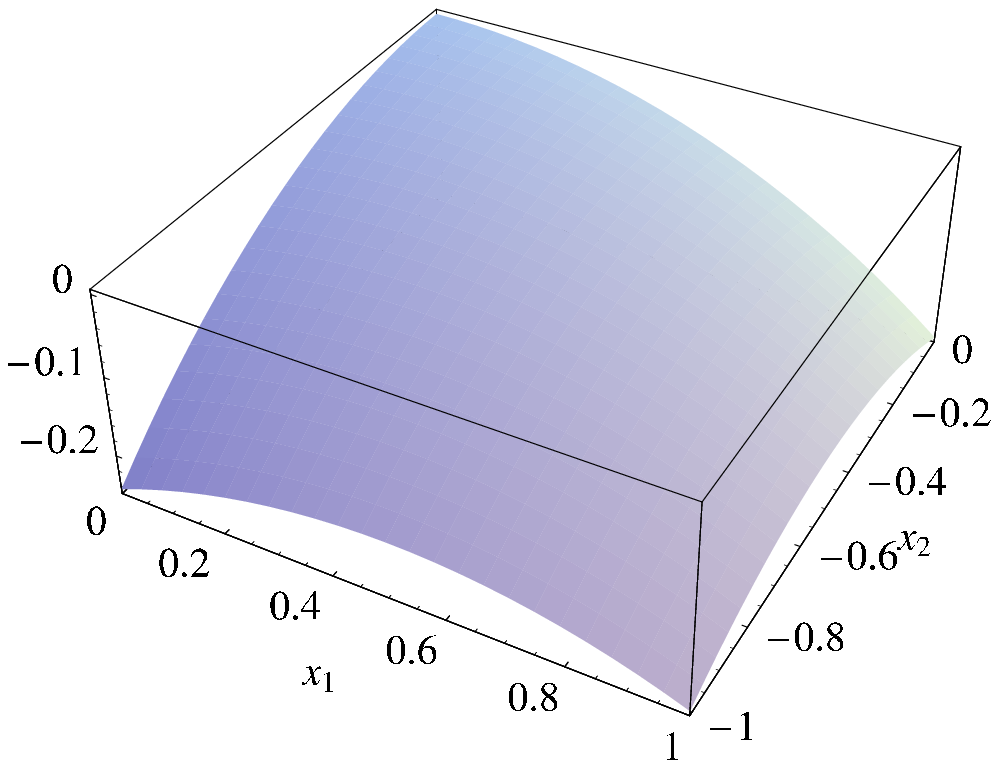,width=2.6in}
\caption{Top left: The canonical symplectic potential $\Gcan$. 
Top right: The Einstein symplectic potential $\G_{\rm E}=\Gcan+h_{\rm E}$. 
Bottom: $h_{\rm E}$. These are plotted in the range $0\le x_1\le1$, $-1\le x_2\le0$, which is one-third of the hexagon; the values on the rest of the hexagon are determined from these by its $D_6$ symmetry. }
\label{hfinalfig}
}

The final complex and symplectic potentials found by the independent implementations are plotted in Figs.~\ref{complexresults} and \ref{hfinalfig}, along with the respective canonical potentials. The results are plotted in the fundamental domain actually simulated, and one should use the $D_6$ symmetry to picture the potentials extended over the whole domain. To compare the \K potential $\F_{\rm E}(u)$ computed in complex coordinates with the symplectic potential $\G_{\rm E}(x)$ computed in symplectic coordinates, we numerically performed a Legendre transform to obtain a \K potential $f'_{\rm E}(u)$ from $\G_{\rm E}(x)$. Plotting $f'_{\rm E}-f_{\rm E}$ in Fig.~\ref{difference}, it can be seen that they differ by less than $5\times 10^{-6}$. (In fact the agreement may be slightly better as there is likely some error introduced in doing the numerical Legendre transform.) Thus our results appear to agree to around the same order that we believe they are accurate.

\FIGURE{
\centerline{\psfig{figure=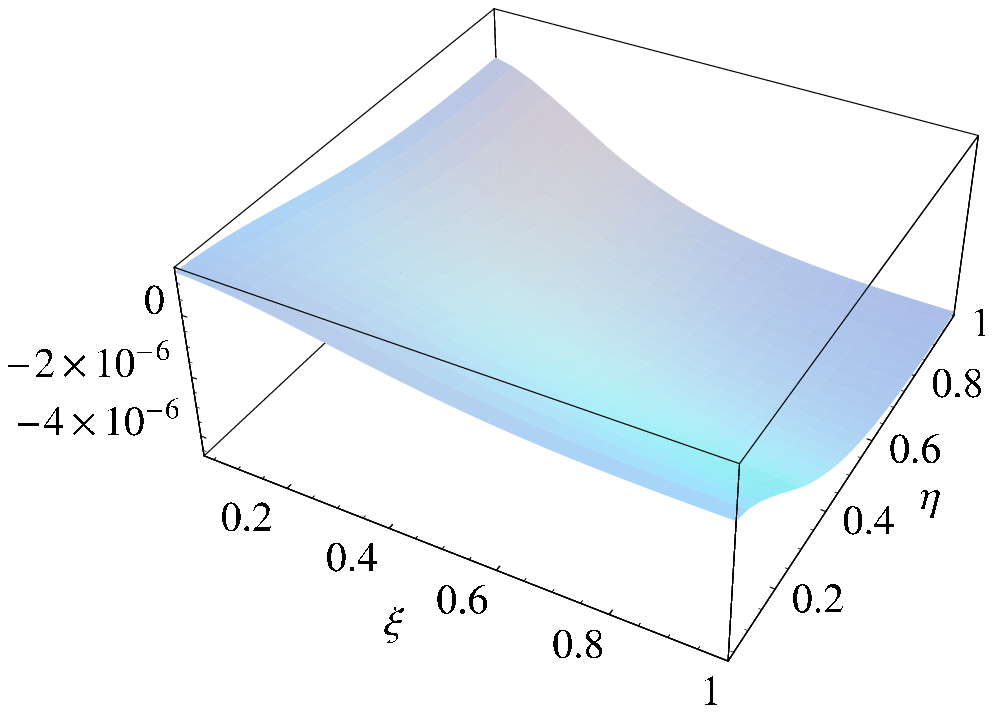, width=3in} }
\caption{The difference between the final \K potential computed in complex coordinates and the Legendre transform of the final symplectic potential computed in symplectic coordinates.
}
\label{difference}
}

\FIGURE{
\epsfig{file=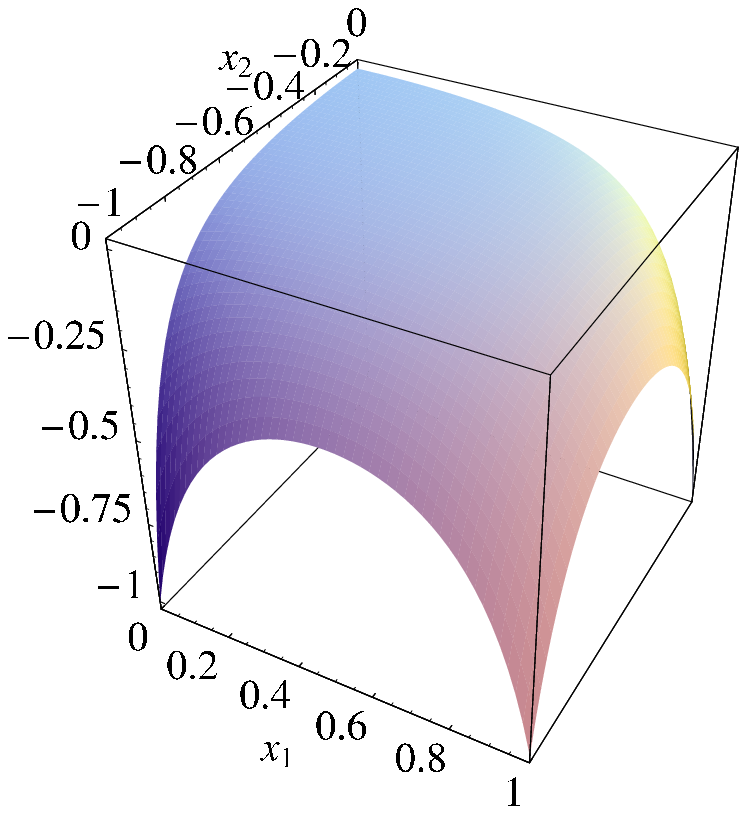,width=2.5in}\quad\epsfig{file=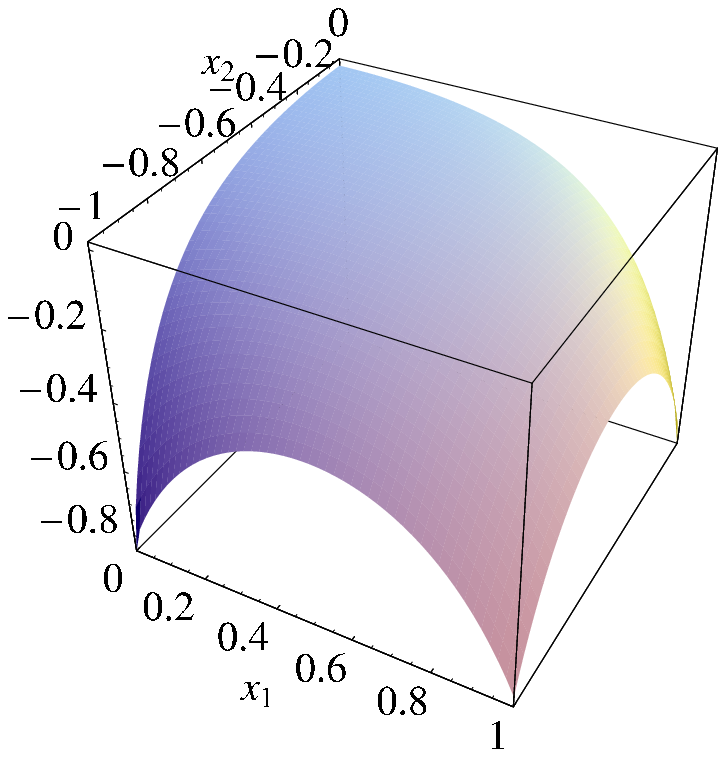,width=2.5in}
\caption{The sectional curvature of the $x_1$-$x_2$ plane at fixed angle $\theta_i$, 
for the canonical (left) and Einstein (right) metric. The two are quantitatively different but qualitatively similar.}
\label{sect}
}

\FIGURE{
\epsfig{file=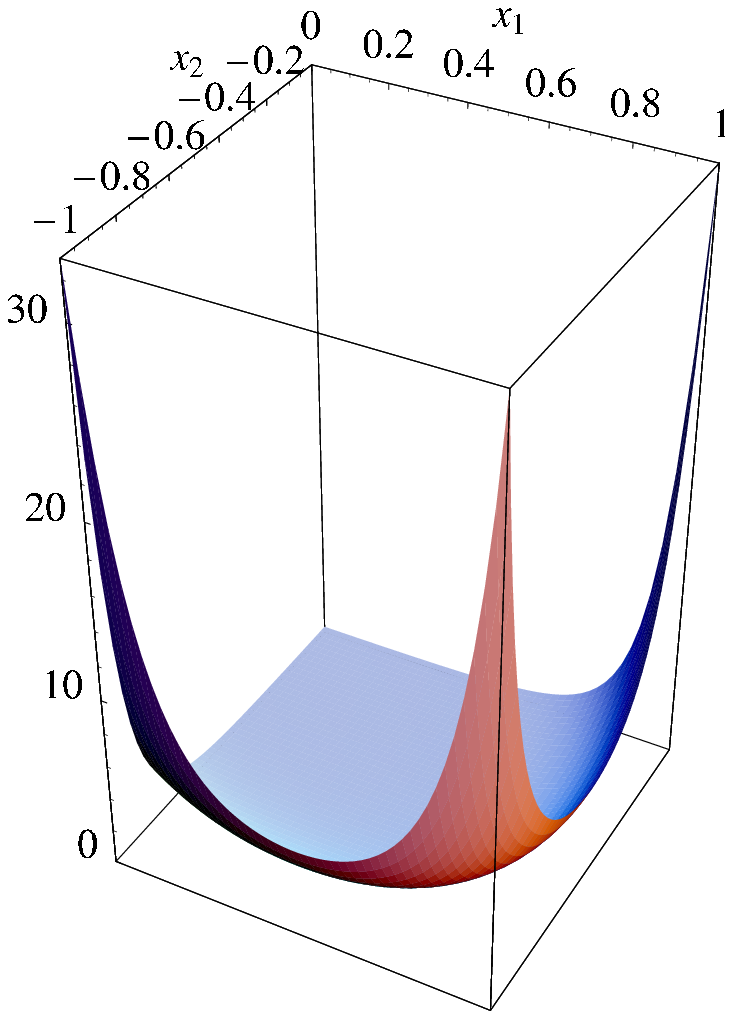,width=2.5in}\quad\epsfig{file=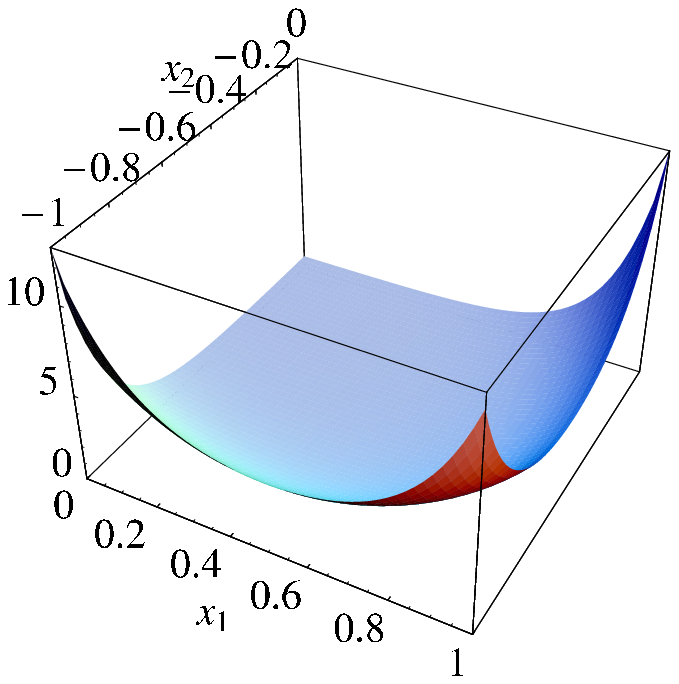,width=2.5in}
\caption{$4\pi^2e$, where $e$ is the Euler density, for the canonical (left) and Einstein (right) metrics. The Euler density is distributed differently but integrates to the same value---2 over this region, and 6 over the whole polytope---for any two metrics. It is more evenly distributed for the Einstein metric than for the canonical metric, but still peaked at the vertices of the polytope.}
\label{Euler}
}

In order to get some feeling for the form of the \K-Einstein metric, and how it compares to the canonical one, it is helpful to plot some curvature invariants. Of course, any invariant depending solely on the Ricci tensor, such as the Ricci scalar, will be trivial, so we need to go to invariants constructed from the Riemann tensor. (Expressions for the Riemann tensor are given in Section \ref{sec:MA} above.) For example, the sectional curvature of the $x_1$-$x_2$ plane (at fixed $\theta_i$), which is $R_{x_1x_2x_1x_2}/\det(G_{ij})$, is plotted for the canonical and Einstein metrics in Fig.~\ref{sect}. Also of interest is the Euler density
\begin{equation}
e = \frac1{32\pi^2}\left(
R^2 - 4R_{\mu\nu}R^{\mu\nu} + R_{\mu\nu\rho\lambda}R^{\mu\nu\rho\lambda}
\right),
\end{equation}
which integrates to the Euler character of the manifold, which is 6 for $dP_3$. The first two terms inside the parentheses cancel in the case of an Einstein metric. Fig.~\ref{Euler} shows $4\pi^2 e$ for the canonical and Einstein metrics. The factor of $4\pi^2$ takes account of the coordinate volume of the fiber. Recalling that $\sqrt{g}=1$ in symplectic coordinates, the plotted quantity should integrate to 2 over the plotted region, which covers one-third of the polytope. This can easily be checked in both cases by numerical integration.

\subsection{Laplacian eigenvalues}

An important geometric quantity is the spectrum of the scalar Laplacian. Here we illustrate a simple method to compute low-lying eigenfunctions. The natural flow associated with the scalar Laplacian is diffusion, and the late time asymptotic behavior of the diffusion flow is dominated by the eigenfunction with lowest eigenvalue. Hence simulating diffusive flow on the $dP_3$ geometry found, and extracting the asymptotics of this flow allows the lowest eigenfunction to be studied. We may classify the eigenfunctions under action of the $D_6$ and $U(1)^2$ isometries. Since the flow equation is invariant under these symmetries, if we start with initial data that transforms in a particular representation, the function at any later time in the flow will remain in this representation. For simplicity we will focus on eigenfunctions which transform trivially, but obviously the method straightforwardly generalizes to compute the low-lying eigenfunctions in other sectors. As for the Ricci flow, the flow does not depend on second normal derivatives of $\psi$ at the boundaries of the hexagon domain, and hence we do not require boundary conditions for $\psi$ there, except to require it to be smooth.

The lowest eigenfunction of $-\lap_{\rm E}$ in the symmetry sector we study 
is $\psi=\mbox{constant}$ which has 
zero eigenvalue. We are interested in the next lowest mode which has 
positive eigenvalue and non-trivial eigenfunction, denoted $\psi_1(x)$ 
with eigenvalue $\lambda_1$. Then we consider the diffusion flow on our del Pezzo solution,
\begin{equation}
\pd{}{t} \psi(t,x) = \lap_{\mathrm{E}} \psi(t,x) 
\end{equation}
and start with initial data for $\psi$ that is symmetric and will hence remain symmetric. At late times,  the flow will generically behave as,
\begin{equation}
\psi(t,x) = \psi_0 + \psi_1(x) e^{-\lambda_1 t} + O(e^{-\lambda_2 t})
\end{equation}
where $\psi_0$ is a constant, corresponding to the trivial zero eigenmode, and $\lambda_2$ is the next lowest  eigenvalue $\lambda_2>\lambda_1$. Waiting long enough and subtracting out the trivial constant, the late flow is given by $\psi_1$, the eigenfunction we wish to compute.

\FIGURE{
\epsfig{file=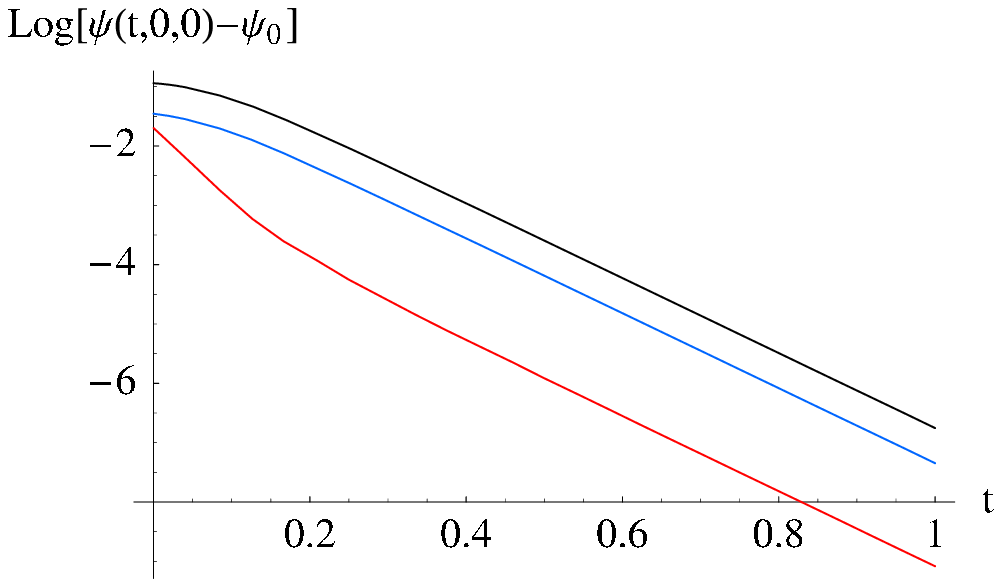,width=3.5in}
\caption{Decay of $\psi$ under diffusion towards a constant as a function of diffusion time $t$. The log 
of $\psi(t,0,0) - \psi_0$ is plotted, the slope giving the eigenvalue for the lowest (non-constant) symmetric eigenfunction. The 3 curves correspond to 3 different initial profiles for $\psi$ although we see that the decay quickly becomes dominated by the lowest eigenmode.}
\label{diffusion}
}

\FIGURE{
\centerline{\epsfig{file=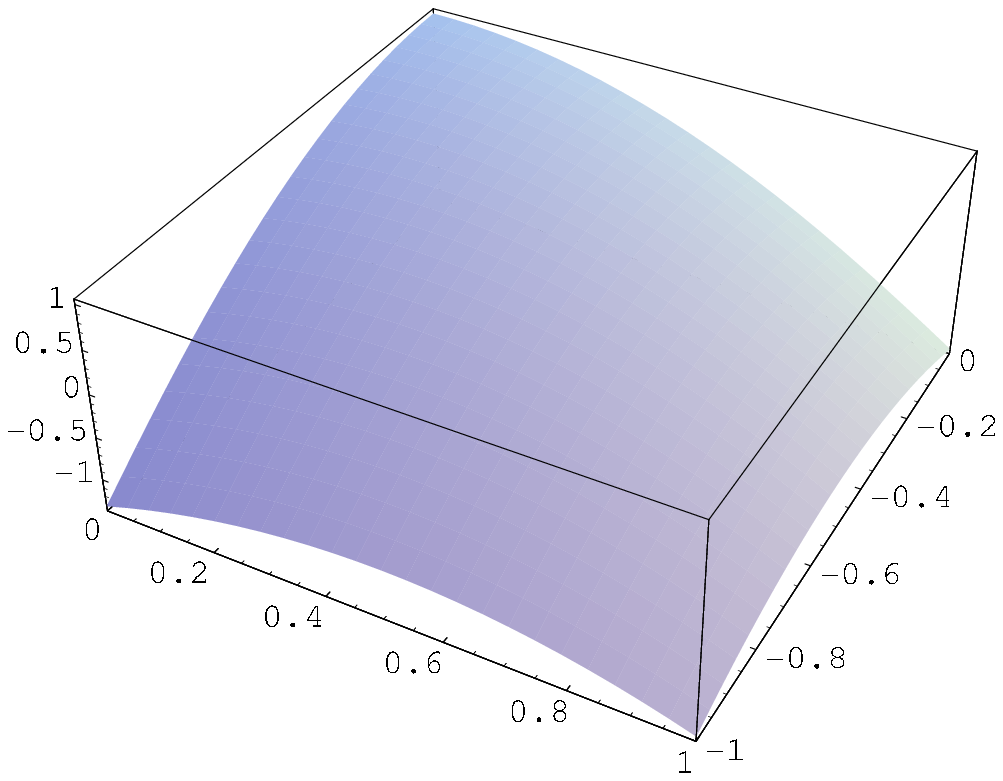,width=3in}}
\caption{The lowest eigenfunction of the scalar Laplacian which transforms trivially under the $D_6$ and $U(1)^2$ symmetries. }
\label{efunction}
}

In Fig.~\ref{diffusion} we plot the log of $\psi(t,0,0)-\psi_0$ as a function of the flow time $t$ for 3 different initial data. Once the higher eigenmodes have decayed away, we clearly see the flows tend to the same exponential behaviour. We estimate this eigenvalue by fitting the exponential decay as $\lambda_1 = 6.32$. In Fig.~\ref{efunction} we plot the eigenfunction $\psi_1(x)$, normalized so that $\psi_1(0,0)=1$. Note that, as expected, for different initial data we consistently obtain the same function. 

This lowest eigenvalue and eigenfunction can also be obtained from the approach to the fixed point of the Ricci flow. For concreteness let us work in symplectic coordinates; corresponding expressions will hold in complex coordinates. Expanding $h$ about its fixed-point value,
\begin{equation}
h = h_{\rm E}+\delta h,
\end{equation}
the flow equation \eqref{heom} becomes, to first order in $\delta h$, 
\begin{equation}
\pd{\delta h}{t} = (\lap_{\rm E}+2)\delta h - \delta c,
\end{equation}
where $\delta c$ depends on how $c$ is chosen. Since we used initial conditions for the Ricci flow that respected the $D_6$ symmetry, we should find that the potentials approach their fixed point values the same way as $\psi$ above, with a shift of 2 in the exponent:
\begin{equation}\label{happraoch}
h(t,x) = h_{\rm E}(x) + \left(\psi_1(x)-\psi_1(0)\right) e^{-(\lambda_1-2) t} + \cdots;
\end{equation}
the eigenfunction is shifted by a constant because $c$ was chosen to keep $h(t,0)=0$ along the flow. Our numerical flows confirm this expectation. The corrections in \eqref{happraoch} involve both higher eigenvalues of the Laplacian and higher-order effects in $\delta h$.

\section{Symplectic polynomials}
\label{sec:polynomial}

In the Ricci flow simulation in symplectic coordinates discussed in Section \ref{sec:symplectic}, the function $h(x)$ --- which encodes the symplectic potential and therefore the metric --- was represented by its values on a lattice of points in $x_1,x_2$. In this section we will discuss a different way to represent the same function, namely as a polynomial in $x_1,x_2$. Since the solution $h_{\rm E}$ to the Monge-Amp\`ere equation is a smooth function, it can be represented to good accuracy with a vastly smaller amount of data in this way: a few polynomial coefficients as compared to values on thousands of lattice points. Furthermore, quite independent of the solutions found in the previous section, the problem of finding an approximate solution to the Monge-Amp\`ere equation can be expressed as an optimization problem for the polynomial coefficients; we will use this fact to develop a third algorithm in the following section that is quite different in character from Ricci flow.

It is interesting to note that the metrics obtained from polynomial expressions for $h(x)$ are the symplectic analogues of the so-called ``algebraic" metrics on Calabi-Yau manifolds that have been used for numerical work by Donaldson \cite{Donaldson} and Douglas et al.\ \cite{Douglasetal1,Douglasetal2}. The algebraic metrics, which are defined for a Calabi-Yau embedded in a projective space, have a \K potential that differs from the induced Fubini-Study one by (the logarithm of) a finite linear combination of a certain basis of functions, namely the pull-backs of the Laplacian eigenfunctions on the embedding projective space. This is a generalization of the usual strategy of representing a function by expanding it in a basis of Laplacian eigenfunctions (such as Fourier modes); since the eigenfunctions on the Calabi-Yau depend on the metric that one is trying to find, one instead uses the eigenfunctions on the embedding space, which are known in closed form (and are indeed very simple). In our case, we consider the embedding of $dP_3$ in $(\CP^1)^3$, which, as discussed in Section \ref{sec:details}, is described in symplectic coordinates by the equation $x_1+x_2+x_3=0$ (where the $x_i$ are the symplectic coordinates on the respective $\CP^1$ factors). %The harmonic functions on $(\CP^1)^3$ that are invariant under its $U(1)^3$ toric isometry group are simply the monomials in $x_1,x_2,x_3$; 
The first $n$ eigenspaces of the Laplacian on $(\CP^1)^3$ (with respect to the Fubini-Study metric), restricted functions that are invariant under the $U(1)^3$ isometry group, are spanned by the monomials in $x_1,x_2,x_3$ up to order $n$. This is precisely the basis of functions we use to expand the difference $h$ between the symplectic potential $g$ and the induced one $\Gcan$.

We now describe some fits to the numerical solutions of the last section in terms of polynomials up to sixth order in $x_1$ and $x_2$, and quantify how well those polynomials do in solving the Einstein equation. These small polynomials likely provide sufficiently accurate approximations to the Einstein symplectic potential for most purposes, while at the same time being more tractable than the full numerical data for analytic calculations.

We begin by noting that, since $h_{\rm E}$ is invariant under the hexagon's $D_6$ symmetry group, it is sufficient to consider invariant polynomials. As shown in Appendix \ref{app:poly}, every invariant polynomial can be expressed in terms of the two basic invariant polynomials,
\be\label{UVdef}
U = x_1^2 + x_1 x_2 + x_2^2 \; , \; \; \; V = x_1^2 x_2^2 (x_1+x_2)^2 \ .
\ee
We simply do a least-squares fit of the polynomial coefficients to the lattice values of $h_{\rm E}$ obtained in the Ricci flow in symplectic coordinates, that is, we minimize
\begin{equation}
\alpha^2 = 
\frac1{V_{dP_3}}\int_{dP_3}\sqrt{g}\left(h_{\rm fit}-h_{\rm E}\right)^2
-\left(\frac1{V_{dP_3}}\int_{dP_3}\sqrt{g}\left(h_{\rm fit}-h_{\rm E}\right)\right)^2 \ .
\end{equation}
(Any constant difference between $h_{\rm fit}$ and $h_{\rm E}$ is irrelevant). 
At successive orders in $x$ we find the following fits:
\begin{equation}
\begin{array}{l|l|l}
h_{\rm fit} & \alpha & \beta \\
\hline 
0	& 0.06 & 0.5 \\
-0.24U & 10^{-3} &0.1 \\
-0.2214U -0.0215U^2 &10^{-4} & 0.03 \\
-0.22412U -0.01450U^2 -0.00521U^3 +0.00734V  & 10^{-5} & 0.007
\end{array}
\label{sympfit}
\end{equation}
In each case we have written only the significant digits of the coefficients.\footnote{These digits do not change between the run with 200 lattice points and the run with 400 lattice points (except the last digit of each coefficient in the sixth-order approximation), and are therefore presumably equal to their continuum values.} Independently of our numerical result $h_{\rm E}$, it is useful to know how far the metric corresponding to $h_{\rm fit}$ deviates from being Einstein. In Fig.~\ref{einstapprox}, the pointwise rms deviation of the eigenvalues of the Ricci tensor from 1,
\begin{equation}
D \equiv \sqrt{\frac14(R_{\mu\nu}-g_{\mu\nu})^2},
\end{equation}
is plotted for these four functions. The global rms deviation from being Einstein, $\beta$, where
\begin{equation}
\beta^2 = \frac1{V_{dP_3}}\int_{dP_3}\sqrt{g}D^2,
\end{equation}
is also shown in the table above. As expected, each successive order gives a substantially better approximation, and a metric that is substantially closer to being Einstein.
\FIGURE{
\epsfig{file=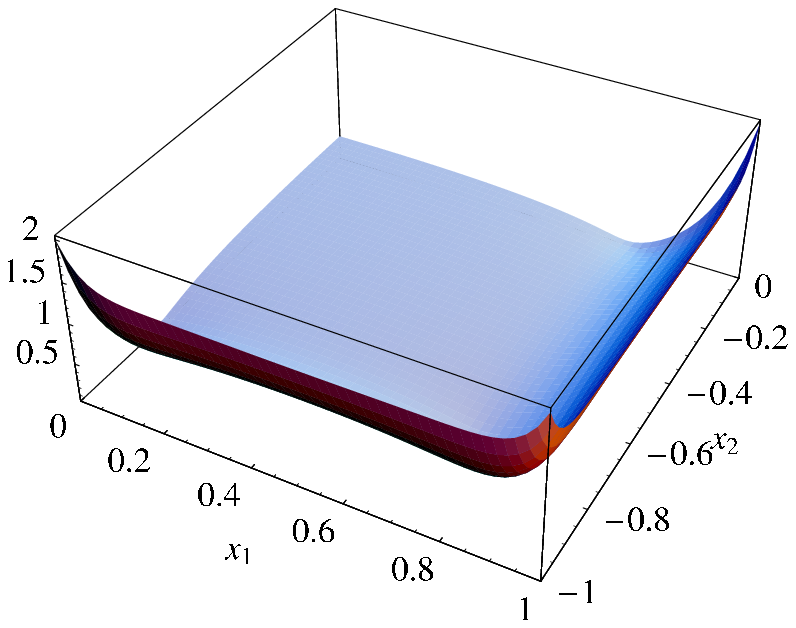,width=3in}\epsfig{file=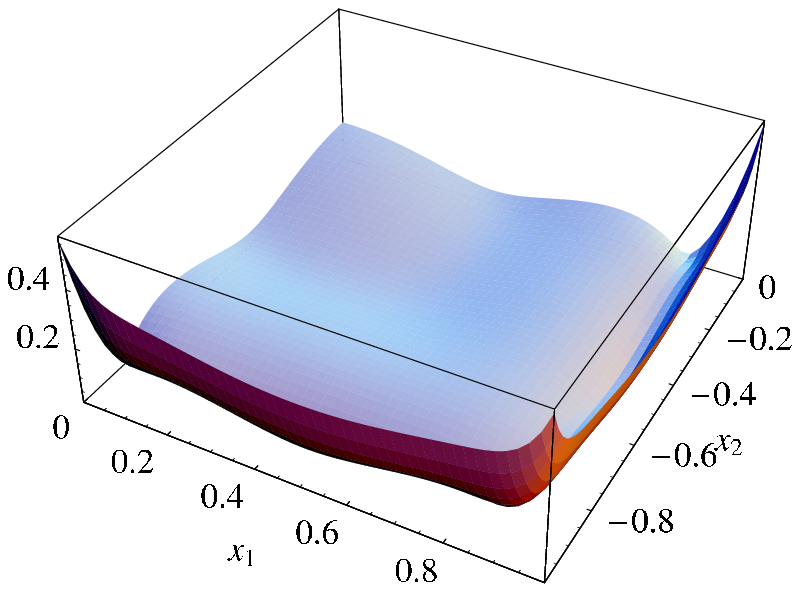,width=3in} \\
\epsfig{file=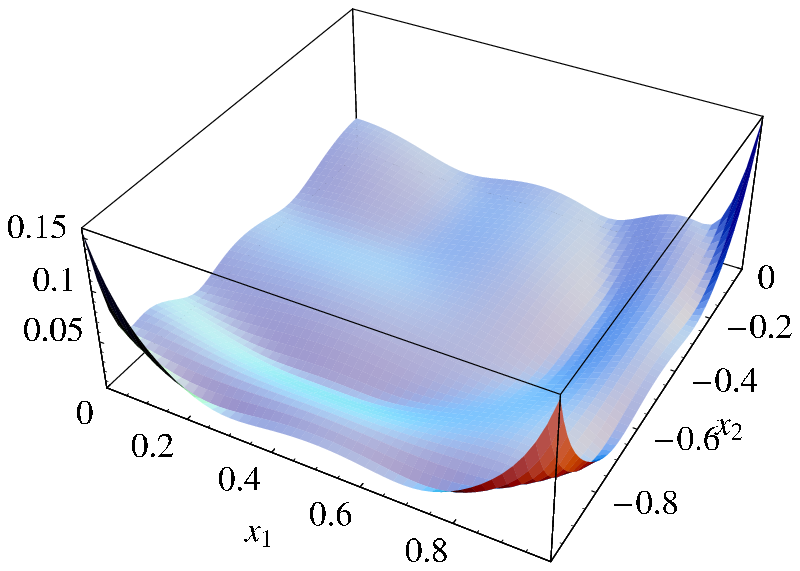,width=3in}\epsfig{file=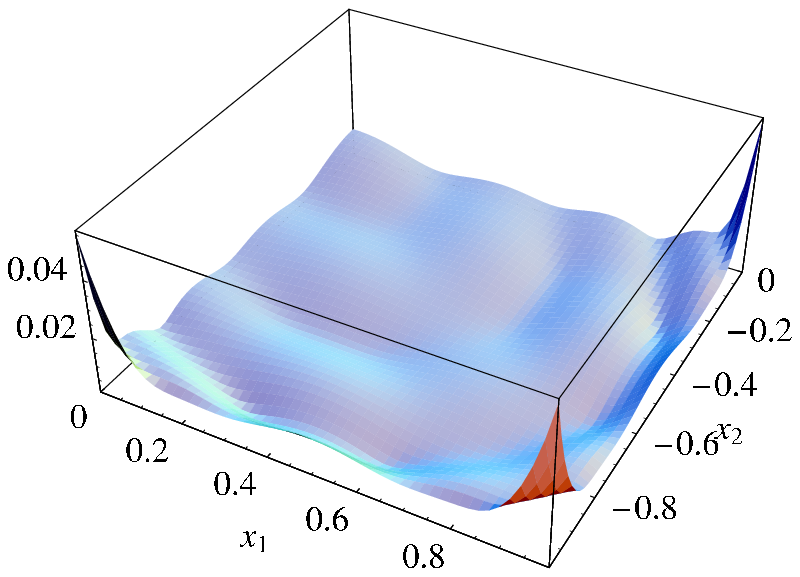,width=3in}
\caption{$D\equiv\sqrt{\frac14(R_{\mu\nu}-g_{\mu\nu})^2}$ versus $x_1,x_2$ for the four polynomial functions listed in \eqref{sympfit}. The deviation from being Einstein is most significant along the edges of the hexagon in the first two cases, and at the corners in the second two.}
\label{einstapprox}
}

For eigenfunctions of the Laplacian, we may perform the same invariant polynomial fits as we did for $h$; at quadratic order we find $\psi_1 \approx 0.985 - 2.37 U $.

We regard table \ref{sympfit} as a key result of this paper. The last line of the table provides in an extremely compact, analytic form, an approximation to the true symplectic potential on $dP_3$. It deviates from the true potential pointwise by at most $\sim 0.1 \%$, and satisfies the Einstein condition well within the hexagon, giving at most a $10\%$ error near the hexagon corners as measured by the pointwise rms deviation of the Ricci tensor eigenvalues, defined above.

\section{Constrained optimization}
\label{sec:leastsquares}

In the previous section we used the results of Ricci flow to find a polynomial approximation to $h(x)$, the smooth part of the \K-Einstein symplectic potential (recall that $h\equiv g-g_{\rm can}$). Here we instead search for a polynomial approximation to $h$ using the Monge-Amp\`ere equation directly. A simple approach
would be to consider the space of polynomials of some given order, 
and minimize an error function built from the Monge-Amp\`ere equation on that space. However, if one wishes to obtain high accuracies, one needs to go to high orders, and then this brute force approach rapidly becomes intractable due to the large number of polynomial coefficients and the difficulty of searching in a high dimensional space. 
We therefore take advantage of the analytic properties of the Monge-Amp\`ere equation to constrain the polynomial coefficients, by requiring the polynomial to solve it order by order in $x_i$. As we will see, this leaves only a small number of undetermined parameters, dramatically simplifying the error function minimization. 
We now explain the details of the method.

We begin by noting that the exact solution $h_{\rm E}(x)$ to the Monge-Amp\`ere equation is an analytic function of the $x_i$, which  
can be seen in a couple of ways. As noted, the symplectic coordinates are eigenfunctions 
of the Laplacian. Hence, in these coordinates the Ricci curvature operator has the 
same character as it does in  harmonic cooordinates: it is actually an elliptic operator. 
Because the Einstein equations are analytic in the metric, they will have analytic solutions. 
Somewhat more directly, the Monge-Amp\`ere equation we are solving is elliptic at the \K-Einstein
potential and analytic, and so the solution will be analytic.  

Before we constrain the polynomial approximation to $h(x)$ using the Monge-Amp\`ere equation, we constrain it by imposing the hexagon's $D_6$ discrete symmetry group. As discussed in Section \ref{sec:polynomial}, any invariant polynomial in $x_i$ can be written in the form
\be
h = \sum_{i,j} c_{i,j} U^i V^j \  ,
\label{powerseries}
\ee
where $U$ and $V$ are given in Eq.\ \eqref{UVdef}. 
To eighteenth order in $x_1$ and $x_2$ we write the series as follows
\begin{eqnarray}
h &=& A_0 + A_1 U + A_2 U^2 + A_3 U^3 + A_4 U^4 + A_5 U^5 
+ A_6 U^6 + A_7 U^7 + A_8 U^8 + A_9 U^9 + \ldots \nonumber \\ 
 & & {+}\, V \, (B_0 + B_1 U + B_2 U^2 + B_3 U^3 
 +B_4 U^4 + B_5 U^5 + B_6 U^6 + \ldots ) \nonumber\\
 & & {+}\, V^2 \, (C_0 + C_1 U + C_2 U^2 + C_3 U^3+\ldots )  + 
 V^3 \, (D_0 + \ldots) + \ldots \ .
 \label{eq:18order}
\end{eqnarray}

Plugging this series into the Monge-Amp\`ere equation \eqref{hequation} (with $\gamma=0$ and $\Lambda=1$), %near the origin of the hexagon 
yields constraints on the $c_{i,j}$ that relate the $c_{i,j}$ with $j>0$ to the $c_{i,0}$. To make the expressions a little simpler, we introduce a new 
constant $\alpha$:
\be
A_0 = - \frac{1}{2} \ln 3 - \ln \alpha \ .
\ee
We worked out the relations up to order 18 in $x_1$ and $x_2$.  
The first relation is that $A_1 = -1 \pm \alpha$.  The numerical Ricci flow
results are consistent only with the plus sign.
The next few relations are
\[
A_2 = -\frac{1}{6} +  \frac{\alpha^2}{4} \; , \;\;\;
A_3 =-\frac{2}{27} -\frac{2 B_0}{27} + \frac{11 \alpha^3}{72} \; ; \;\;\; \\
A_4 = -\frac{1}{28} - \frac{5\alpha}{378} -\frac{25}{189} \alpha B_0 + \frac{145 \alpha^4}{1152} \ ; \\
\]
\[
B_1 = \frac{25}{14} \alpha B_0 + \frac{1}{28} (-4 + 5 \alpha) \; ; \;\;\; 
B_2 =  \frac{85}{32} \alpha^2 B_0 + \frac{1}{192} (-32 + 51 \alpha^2)  \ .
\]
%\begin{eqnarray*}
%A_2 &=& -\frac{1}{6} +  \frac{\alpha^2}{4} \ , \\
%A_3 &=&-\frac{2}{27} -\frac{2 B_0}{27} + \frac{11 \alpha^3}{72} \ , \\
%A_4 &=& -\frac{1}{28} - \frac{5\alpha}{378} -\frac{25}{189} \alpha B_0 + \frac{145 \alpha^4}{1152} \ , \\\
%B_1 &=& \frac{25}{14} \alpha B_0 + \frac{1}{28} (-4 + 5 \alpha) \ , \\
%B_2 &=&  \frac{85}{32} \alpha^2 B_0 + \frac{1}{192} (-32 + 51 \alpha^2)  \ .
%\end{eqnarray*}

The power of this method is that since the Monge-Amp\`ere equation determines many of the coefficients, when we determine the remaining ones by minimizing an error function, then at a given order there are far fewer parameters to solve for.
Truncating at eighteenth order in $x_1$ and $x_2$, we need fit only 4 parameters $A_0$, $B_0$, $C_0$, and $D_0$ to find an approximation to $h$. In contrast, minimizing an error function using the most general unconstrained eighteenth order polynomial given in equation \ref{eq:18order} involves searching a 22 dimensional space.

In principle if we took an arbitrarily high order expansion, then about the origin of the hexagon, in the region where the series for $h$ converges, we would solve the Monge-Amp\`ere equation precisely. However we see that we still have undetermined constants in the series, and these correspond to the fact that %in an elliptic PDE 
we must provide %the correct 
boundary conditions to determine a solution fully.

A posteriori, our numerics strongly suggest that $h$ converges everywhere in the interior of the Delzant polytope.
%
%Whilst we have not been able to prove that this 
%expansion about the origin of the smooth part of 
%the symplectic potential, $h$, converges everywhere 
%within the Delzant polytope, we will assume here that it does so.
Constraining $h$ to have the correct behavior on the boundary of the polytope should completely determine the remaining constants in the power series expansion. 
%We will see that although this is taken on assumption, 
%our numerical results provide strong evidence that this is indeed the case.
%
Hence we fit the remaining parameters by requiring Eq.~(\ref{hequation}) be satisfied at the boundary of the hexagon. %(with $\Lambda=1$ and $\gamma=0$).  
As emphasized in the discussion around Eq.~(\ref{BCcpone}), the boundary conditions do not need to be supplied separately --- they are enforced by Eq.~(\ref{GPDE}) itself.  For $dP_3$, along $x_1=1$, Eq.~(\ref{GPDE}) reduces to the boundary condition
\be
{\mathcal BC} \equiv 1-2x_2 -2x_2^2 + (1-x_2^2) (1-(1+x_2)^2)) h_{x_2x_2} 
- \exp \left[2 (h_{x_1} + x_2 h_{x_2} - h)\right] = 0 \ .
\label{mybc}
\ee
To find this expression, we have assumed that $h$ is smooth at the boundary $x_1=1$. At the corners $x_2 = 0$ and $x_2=-1$, Eq.~(\ref{mybc}) reduces further to 
$h= h_{x_1}$ and $h=-h_{x_2} + h_{x_1}$ respectively. Explicitly, we determine 
the remaining parameters ($A_0$ through $D_0$ in the eighteenth order truncation) by minimizing
\be
M = \sum_p |{\mathcal BC}(p)|^2 \ ,
\ee
summed over twenty equally spaced points along the boundary $x_1=1$.  
We find, as we include more terms in the series,
 \be
 \begin{array}{c|c|c|c|c|c}
 \mbox{order} & M & \alpha & B_0 & C_0 & D_0 \\ \hline
 2 & 0.02 & 0.757 &&\\
 %4 & 0.003 & 0.7726 &&\\
 6 & 9 \times 10^{-5} & 0.7753 & 0.011&\\
 %8 & 2 \times 10^{-5} & 0.7760 & 0.0059 &\\
 10 & 3 \times 10^{-6} & 0.77616 & 0.00508& \\ 
 %12 & 3 \times 10^{-7} & 0.776209 & 0.00486 & -0.003 &\\
 14 & 5 \times 10^{-8} & 0.776226 & 0.00480 & -0.00055 &\\
 %16 & 1 \times 10^{-8} & 0.776233 & 0.004787 & -0.00023&\\
 18 & 2 \times 10^{-9} & 0.776235 & 0.004781 & -0.00015 & 0.004 
 \end{array}
 \label{leastsquares}
 \ee 
The full expression for the 18th-order polynomial is given in a Mathematica notebook available for download at the websites \cite{website}.
In Appendix \ref{app:convfit} we show the dependence of $h$ at various locations in the polytope as a function of the number of terms taken in the expansion. We see that convergence for $h$ is fast --- apparently faster than polynomial --- in the number of terms, and in particular for everywhere tested within the hexagon, and also on its boundary, we see convergence. In particular we see no sign of poor behaviour near or on the boundaries of the hexagon. We also see that the values the series converges to are in excellent agreement with the continuum extrapolated values of $h$ found from the Ricci flow method and detailed in Appendix \ref{sec:numerics}. From these data we estimate that the potential given by the eighteenth order expansion differs from the true solution by approximately one part in $10^6$, and hence is comparable in this respect to the $400\times400$ Ricci flow result.

The figure of merit $M$ does not give a very good indication of the degree of 
accuracy of our fit globally.  To understand how well we are doing globally, 
we use the same local estimate of error as in the previous section, $D = \sqrt{\frac14(R_{\mu\nu}-g_{\mu\nu})^2}$. We find that the maximum value $D$ attains in the domain decreases with each increase in order of the expansion. The maximum is found on the lines connecting the origin to the hexagon vertices, and hence in Fig.~\ref{fittingerror}, we plot this error estimate along one of these lines, $D(x_1,0)$, for the 6th, 12th and 18th order polynomial expansions. As expected the error is smallest at the origin, and most error is localized near the boundaries. Since the error is rather localized near the hexagon vertex, we have plotted this error against $\ln{(1-x_1)}$ to demonstrate that it is indeed finite at the vertex. We see that while the symplectic potential taken pointwise may be accurate to 1 part in $10^6$ as stated above, since the error is localized in the hexagon corners, the quality of the solution is a little worse in these regions. We see for the 18th order approximation that the error in the Einstein condition, estimated by $D$, is about one part in $10^3$ at the vertex.

\FIGURE{
\centerline{\epsfig{file=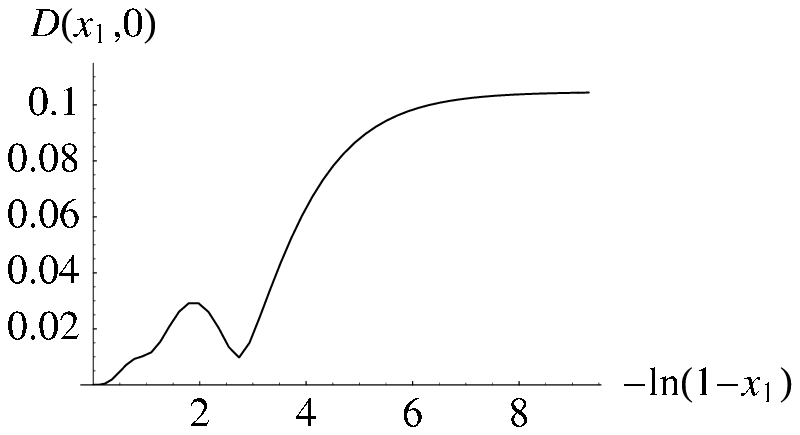, width=2.5in}\epsfig{file=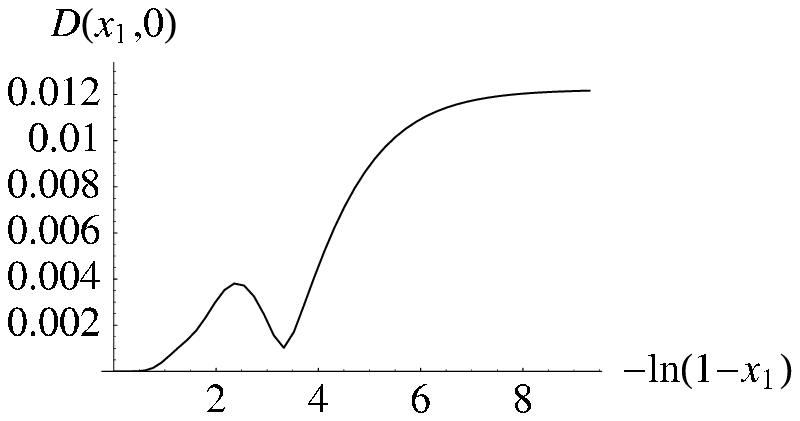, width=2.5in}\epsfig{file=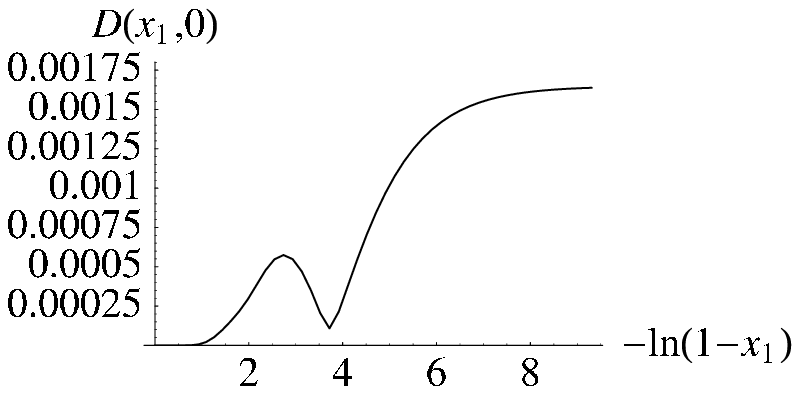, width=2.5in}}
\caption{$D$, the error in the Einstein condition, along the line $x_2=0$ from the origin $x_1=0$ to the hexagon corner $x_1=1$ where the maximum of $D$ in the hexagon domain occurs. The left plot is for a 6th order expansion, the middle is 12th order, and the right is 18th order. We see the error in the Einstein condition is quite localized near the hexagon corner (hence we plot against $\ln{(1-x_1)}$), but remains finite there, and decreases everywhere with increasing numbers of terms in the expansion.}
\label{fittingerror}
}

\subsection{Laplacian eigenvalues}

We can use this 18th order fit to extract the lowest eigenvalues (of eigenfunctions invariant under the $D_6$ action) on our manifold.  Note that 
in coordinates where the metric is analytic, eigenfunctions of the Laplacian are analytic: 
elliptic equations with analytic coefficients have analytic solutions. Thus we can play a
very similar game, expressing the eigenvectors as a local power series
near the origin of the hexagon in the $U$ and $V$ variables:
\be
\psi(U,V) = \frac{1}{10} + X_1 U + X_2 U^2 + X_3 U^3 + Y_1 V  + \ldots \ .
\ee
We have chosen to normalize $\psi(0,0) = 0.1$.  
We could in principle have used the differential equation to constrain some of the $X_i$ and $Y_i$, but
we did not, aiming for a fit whose errors are more evenly distributed over the hexagon. 
We fit a hundred equally spaced points in a square domain
$0< x_1 < 0.9$ and $-0.9 < x_2 < 0$ and minimize
\be
M_\psi = \sum_p |\lap \psi + \lambda \psi|^2 \ ,
\ee
as a function of $\lambda$ and the $X_i$ and $Y_1$.
By searching for successive local minima of $M_\psi$, we can extract successively higher
eigenvalues.  We find
\be
\begin{array}{c|c|c|c|c|c|c}
\mbox{order} & M_\psi & X_1 & X_2 & X_3 & Y_1 & \lambda_1 \\ \hline
2 & 0.006 & -0.239 & & & & 6.27 \\
4 & 0.002 & -0.246 & 0.011 & & & 6.325 \\
6 & 3 \times 10^{-5} & -0.245 & 0.006 & 0.006 & -0.024 & 6.322
\end{array}
\ee
\be
\begin{array}{c|c|c|c|c|c|c}
\mbox{order} & M_\psi & X_1 & X_2 & X_3 & Y_1 & \lambda_2 \\ \hline
4 & 0.5 & -0.69 & 0.79 & & & 17.4 \\
6 & 0.008 & -0.67 & 0.70 & 0.16 & -1.28 & 17.2
\end{array}
\ee
For the lower eigenvalue $6.32$, note that the ratio of the first two coefficients $-2.39$
is in reasonably good agreement with the corresponding ratio 
$-2.37/0.985 = -2.41$ determined in Section 
\ref{sec:polynomial}.

The choice to minimize $M_\psi$ in the domain $0< |x_i| < 0.9$ was a 
compromise that requires
some justification.  First, since (\ref{GPDE}) enforces its own boundary conditions,
minimizing $M_\psi$ close to the boundary $x_1 = 1$ is enforcing the boundary
condition to first order in $x_1-1$.  Second, the power series approximation for
$\psi$ experimentally does not appear to have good convergence properties near
the boundary.  Experimentally, the best values for the coefficients of the truncated
power series (in the sense of
agreeing with the coefficients of the power series itself) are obtained
by making a compromise
between minimizing over a set of points that extends to the boundary and minimizing
over a set of points for which the power series has good convergence properties.

\subsection{Harmonic (1,1)-forms}

As noted in Section \ref{harmonicforms}, 
harmonic $(1,1)$-forms and eigenfunctions of the Laplacian must
satisfy a very similar equation.  We end this section with a computation of the harmonic
$(1,1)$-form $\theta_a$.  Unlike the eigenfunctions computed above, $\theta_a$ does not
transform trivially under $D_6$.  Thus, we assume that $\mu_a$ has a more general
expansion of the form
\be
\mu_a = \ln(1+v_a \cdot x ) + \sum_{n,m} c_{nm} x_1^n x_2^m \ ,
\ee
We use the same least squares approach as above,
minimizing
\be
M_\theta = \sum_p | \lap \mu_a - \mbox{const} |^2 \ .
\ee  
Because of the explicit $x_1 \leftrightarrow x_2$ symmetry, we start with $a=2$
and set $c_{nm}=c_{mn}$.  Fitting
to sixth order in $x_1$ and $x_2$, we find
\be
\begin{array}{lr | lr | lr | lr}
c_{10} =& -0.2250  \\
c_{20} = &0.0638 & c_{11} = &0.0311 \\
c_{30} =  &-0.0301 & c_{21} =  &0.0059 \\
c_{40} = &0.0126 & c_{31} = &0.0005 & c_{22}= &-0.0073 \\
c_{50} = &-0.0150 & c_{41} = &-0.0245 & c_{32} = &-0.0240 \\
c_{60} = &0.0088 & c_{51} = &0.0223 & c_{42} = &0.0281 & c_{33} = &0.0196 \
\end{array}
\ee
The value of $M_\theta \sim 10^{-4}$ at the minimum implies an average
error of $10^{-3}$ at each of the 100 points.  Note the error gets much worse
outside the fitting domain $|x_i| > 0.9$.  The philosophy in this section is similar
to that in the discussion of eigenfunctions:  we are attempting to find more accurate
values of the $c_{ij}$ rather than attempting to minimize the global error.

The fit also yields $F^{ij} \theta_{ij}  = 0.6672$, consistent with our expectations.
We know that 
\be
\omega =  \frac{1}{2} \sum_{a=1}^6 \theta_a \ .
\ee
Clearly $F^{i \bar \jmath} F_{i \bar \jmath} = 2$.  By the dihedral symmetry group,
the value of $F^{i \bar \jmath} \theta_{i \bar \jmath}$ should be independent of
$a$.  We conclude that
\be
(\theta_a)_{i \bar \jmath}  F^{i \bar \jmath} = \frac{2}{3} \ .
\ee

From $\theta_2$, we can reconstruct the other $\theta_a$ by applying the $D_6$ group
action.  To test how good our approximation to $\theta_2$ was, we computed
$\Theta^2 = \Theta^{ij} \Theta_{ij}$ where 
$\Theta_{ij} =  \sum_a (\theta_a)_{ij}$ using our best fit for $\theta_2$.   
Since $\Theta_{ij}$ should be $2 \Fij$, $\Theta^2$ should be approximately 
eight.  A plot of $\Theta^2$ is shown in Fig.~\ref{thetatraceplot}.  

\FIGURE{
\centerline{\epsfig{file=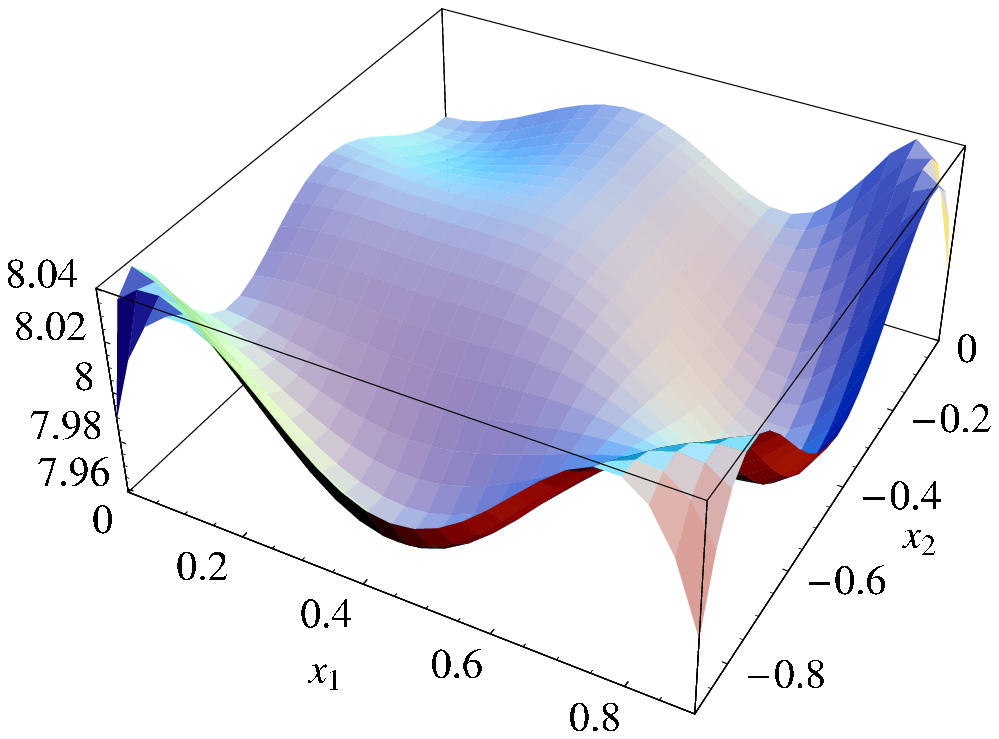, width=3in}}
\caption{The value of $\Theta^2$
for our 6th order fit.}
\label{thetatraceplot}
}

For the purposes of the KT solution described in the Introduction, we need a $\theta$ such
that $\theta^{ij} F_{ij} = 0$.  From the preceding discussion, any linear combination of the form
$\sum_{a} c_a \theta_a$ such that $\sum_a c_a = 0$ will have this property.   We also require
that $\star \theta = -\theta$.  In fact, the condition $\sum_a c_a = 0$ enforces anti-self-duality.  
The reason is that the Hodge star treated as a linear operator acting on 
the space of harmonic (1,1)-forms has signature $(+---)$.  We know that $\star \omega = \omega$; thus
any (1,1)-form orthogonal to $\omega$ must be anti-self-dual.
In general, the numerics suggest that for such a $\theta$, $\theta^{ij} \theta_{ij}$ will be a nontrivial
function of both $x_i$ and thus that solving for $h(p)$ requires solving a PDE in three
real variables.

\section{Discussion}
\label{sec:discussion}

In this paper we have described three different methods to find the \K-Einstein metric on $dP_3$. All three methods exploit the \K and toric structures of the manifold, allowing us, using modest computing resources, to compute the metric in both \K and symplectic coordinates to an accuracy of one part in $10^6$. The results of the different methods are consistent to within that error. We expect that this accuracy is sufficient if one wishes to compute geometric quantities for either physical or mathematical applications, and we have made available the data along with Mathematica notebooks to allow manipulation of these results \cite{website}.

We noted that, for a lesser accuracy of one part in $10^3$, a simple expression for the smooth part $h$ of the symplectic potential, $\G = \Gcan + h$, already provides such an approximation, and we repeat it here:
\begin{eqnarray*}
h(x_1, x_2)  & =&  -0.22412U -0.01450U^2 -0.00521U^3 +0.00734V , \\
U & = & x_1^2 + x_1 x_2 + x_2^2 , \; V = x_1^2 x_2^2 (x_1+x_2)^2 \ .
\end{eqnarray*}
where $\Gcan$ is given in equation \ref{canpotl}. The resulting metric satisfies the Einstein condition everywhere to better than $10\%$ as discussed in Section \ref{sec:polynomial}.

Simulation of Ricci flow has proven to be an effective way to solve the Einstein equation. We have found that implementing the flow is a little simpler in symplectic coordinates: the domain is naturally compact; the symmetries are more manifest; and the boundary conditions are simpler. Our codes (which were not optimized for speed) converged in a few hours for the highest resolutions. For higher accuracy than  attained here, one could optimize the flow simulation, for example by taking more advantage of the discrete symmetries than we have done. More generally, Ricci flow simulation, using an explicit finite differencing method as we have done, can be thought of as a particular iterative scheme for solving the Monge-Amp\`ere equation. If one is interested only in solving that equation, and not in accurately simulating Ricci flow, then this scheme could be modified to improve speed. For example, by replacing the Jacobi-type updating method by a Gauss-Seidel method, one obtains a faster algorithm (experimentally, 50\% faster in complex coordinates). To obtain a parametric improvement in speed would likely require a non-local modification such as multi-grid. 

The constrained optimization approach we have demonstrated uses the symplectic polynomials, reducing the size of the search space by solving the Monge-Amp\`ere equation order by order in $x_i$, and has proven very powerful. It is as accurate as the Ricci flow results, but is quicker. One drawback is that the hexagon origin is singled out as the point where the solution is best, and the error in the solution becomes tightly localized at the corners of the hexagon. Most computational time is invested in determining the constraints in the series expansion, and this algebraic problem gets worse the more terms that are included in the expansion. However, once one has this solution, the numerical minimization of the error function is simple. 

The two methods are complementary in the sense that for Ricci flow the time is spent in numerically computing the flow, whereas for the optimization the time is spent algebraically computing the expansion of the potential. The principle advantage of Ricci flow is that the method is very general, and while it benefits from the \K and toric structures it certainly applies to more general problems which do not possess them. It is not clear how widely applicable the constrained optimization approach is, as it likely works due to the special properties resulting from those mathematical structures. However, it would be interesting to investigate its application to other situations. It would also be interesting to compare these approaches, particularly the constrained optimization, with Donaldson's method \cite{Donaldson}.

\section*{Acknowledgments}
We would like to thank Ken Bube, Simon Donaldson, Richard Hamilton, Julien Keller, Liam McAllister, Gang Tian, and Ursula Whitcher for discussion.
C.H. would like to acknowledge the support of the String Phenomenology Workshop at the KITP, UCSB and the Physics Department at the University of Texas, Austin where part of this work was done.
C.D. is supported in part by a Royalty Research Fund Scholar Award from 
the Office of Research, University of Washington. 
M.H. is supported  by the Stanford Institute for Theoretical Physics and by NSF grant PHY 9870115.
C.H. is supported in part by U.S. Department of Energy under Grant No.~DE-FG02-96ER40956 and by the National Science Foundation under Grants No.~PHY99-07949 and No.~PHY-0455649.
J.K. is supported in part by a VIGRE graduate fellowship.
T.W. is supported by a PPARC advanced fellowship and the Halliday award.

\begin{appendix}

\section{Canonical metric on $\mathbb{CP}^2$}
\label{sec:projective}

Consider $\mathbb{CP}^2$ which has the fan, unique up to
$SL(2, \mathbb Z)$ transformations, $v_1 = (1,0)$, $v_2 = (0,1)$, and $v_3 = (-1,-1)$.
We choose all the $\lambda_a=1$ in $\Gcan$ to assure that the class of the resulting 
K\"ahler form is proportional to the first Chern class,
\be
\Gcan = \frac{1}{2} \Bigl[
(1+x_1) \ln (1+x_1) + (1+x_2) \ln (1+x_2) + (1-x_1 - x_2) \ln (1-x_1 - x_2) \Bigr] \ .
\ee
%\[
%(1-x_1 - x_2) \ln (1-x_1 - x_2) ) \ .
%\]
From $\Gcan$, we can reconstruct the complex coordinates
\be
1 + x_1 = \frac{3 e^{2u_1}}{1+ e^{2 u_1} + e^{2 u_2}} \; ; \; \; \;
1 + x_2 = \frac{3 e^{2 u_2}}{1+ e^{2 u_1} + e^{2 u_2}} \; ;
\ee
\[
1-x_1-x_2 = \frac{3}{1+ e^{2 u_1} + e^{2 u_2}} \ .
\]
The canonical K\"ahler potential is then
\be
\Fcan = \frac{3}{2} \ln (1+ e^{2u_1} + e^{2 u_2} ) - \frac{3}{2} \ln 3 - u_1 - u_2 \ .
\ee
The expression $\Gcan$ satisfies (\ref{GPDE}) provided $c = \ln (4/3)$ and 
$\gamma=0$.
For $\mathbb{CP}^2$, the canonical metric is the Fubini-Study
metric.  In terms of the traditional homogenous coordinates $(X_1, X_2, X_3)$
on $\mathbb{CP}^2$,
the K\"ahler potential is traditionally written, with a different choice of normalization of the volume,
 in the patch $X_3 \neq 0$
\be
\ln (1 + |X_1/X_3|^2 + |X_2/X_3|^2) \ .
\ee
Thus, we identify $|X_1 / X_3| = \exp(u_1)$ and 
$|X_2/X_3| = \exp(u_2)$.
The canonical metric is always K\"ahler-Einstein for Cartesian products of
projective spaces \cite{Guillemin}.

%\section{Details of implementation and error estimates of the numerical Ricci flow}
\section{Numerical Ricci flow implementation and error estimates}
\label{sec:numerics}

In this appendix we give technical details of our two finite difference implementations of Ricci flow, and also discuss the continuum convergence and estimate errors for the resolutions used.

\subsection{Implementation in complex coordinates}

We work on a square domain slightly bigger than a unit polydisk, 
$0 \leq \xi \leq L$ and $0 \leq \eta \leq L$ ($L > 1$)
where $\F$ is assumed to have Neumann boundary conditions along 
$\xi=0$ and $\eta=0$.  Along the internal boundaries, $\xi = L$ and $\eta=L$,
we employ a kind of periodic boundary condition enforced by the $\mathbb{Z}_6$
symmetry of the hexagon.   We take
\be
\F(\xi, L) = \F(1/L, \xi L) + \ln L \ ,
\ee  
to map the boundary $\eta=L$ for $0 \leq \xi  < 1/L$ back into our square
domain.  For $1/L < \xi < 1$, we take a composition of the above map:
\be
\F(\xi, L) = \F( 1/ (L \xi), \xi) + \ln L^2 \xi \ .
\ee
For the $\xi=L$ boundary, we take the preceding rules with $\eta$ and $\xi$ switched.
For the points along the boundary with $\xi>1$ or $\eta>1$, 
we use the rule
\be
\F(\xi,\eta) = \F(1/\xi, 1/\eta) + 2 \ln \xi + 2 \ln \eta \ . 
\ee
Note that as indicated in Section \ref{sec:complex}, due to the $\mathbb{Z}_6$ symmetry, for any point 
$p=(\xi,\eta)$, with $\xi>1$ or $\eta>1$, the value of $\F(p)$ should be related by a K\"ahler transformation to 
the value of $\F$ at a point inside the unit polydisk. 

Although we are only enforcing the $\mathbb{Z}_6$ symmetry along the boundary
of our coordinate patch, the symmetry will hold globally.
First, our initial potential $\Fcan$ respects the symmetry.  Second,
Ricci flow preserves the symmetry. 

To discretize (\ref{feqn}), we approximated $\F(\xi, \eta)$ by its values $\Fd_{IJ}$ on
an $N\times N$ grid with lattice
spacing $\ls = L / (N-1)$ and used standard second order finite differencing for
derivatives. In evaluating the first order derivatives at the boundary, we took
\be
\left. \frac{1}{\xi} \frac{\partial \F}{\partial \xi}\right|_{\xi=0} = 
\left. \frac{\partial^2 \F}{\partial \xi^2} \right|_{\xi=0} 
\; \mbox{and} \;
\left. \frac{1}{\eta} \frac{\partial \F}{\partial \eta}\right|_{\eta=0} = 
\left. \frac{\partial^2 \F}{\partial \eta^2} \right|_{\eta=0} 
\ .
\ee

To impose a discrete version of the boundary conditions, we added extra
rows and columns along the grid.  To impose Neumann boundary conditions,
we imposed that $\Fd_{0,J} = \Fd_{2,J}$ and $\Fd_{I,0} = \Fd_{I,2}$.  To impose the periodic
boundary conditions along the $(I,N)$ and $(N,J)$ boundaries, 
we mapped the the point $(I,N+1)$ or $(N+1,J)$ 
back inside the grid using
the symmetries and used a bicubic interpolation to compute a best
value for $\F$.

\subsection{Implementation in symplectic coordinates}

We represented $h$ on a uniform square lattice in the variables $x_1$, $x_2$. Such a lattice has several advantages, aside from simplicity. First, according to \eqref{sympmetric}, these points are also spread uniformly according to the measure of any symplectic metric. Second, the lattice is itself invariant under the symmetries of the hexagon (since these are elements of $GL(2,\Z)$). In fact, in view of this symmetry group, the lattice should in a sense be considered triangular, with ``edges" running not just horizontally and vertically, but also along the diagonals with slope $-1$. In other words, a given lattice point $(x_1,x_2)$ has six nearest neighbors: ${(x_1\pm\ls,x_2)}$, $(x_1,x_2\pm\ls)$, and $(x_1\pm\ls,x_2\mp\ls)$ (where $\ls$ is the lattice spacing). First and second derivatives were calculated using these nearest neighbor points in a way that was accurate to second order in the lattice spacing and respected the hexagon symmetries. Generally speaking, the lattice spacing was chosen to be one over an integer, so that the polytope boundaries passed through lattice points. In view of the free boundary conditions for $h$, on these boundaries the necessary derivatives were computed by extrapolation, using next-to-nearest neighbor points to give third order accuracy.

\subsection{Simulation of Ricci flow}

The Ricci flow was simulated by an explicit method, with first-order accurate time derivatives and a time step of $\propto \ls^2$. The constant of proportionality is of order one, but depends on the initial metric since the equation is non-linear. For the canonical choices of initial potentials in the complex coordinate case we required a time step $\frac17 \ls^2$, while in the symplectic case we required $\frac12 \ls^2$. However modifying the initial potential may require a smaller initial timestep.

The diffusive nature of the flow requires the time step to be the square of the spatial lattice interval. The errors in the time derivatives are therefore of the same order as in the spatial derivatives. In order to proceed to much higher resolutions implicit differencing, such as the Crank-Nicholson scheme should be used, or the time steps would become prohibitively small. However, for the resolutions we have used here, the explicit method is quite manageable.

\subsection{Convergence tests}

\FIGURE{
\centerline{a) \epsfig{file=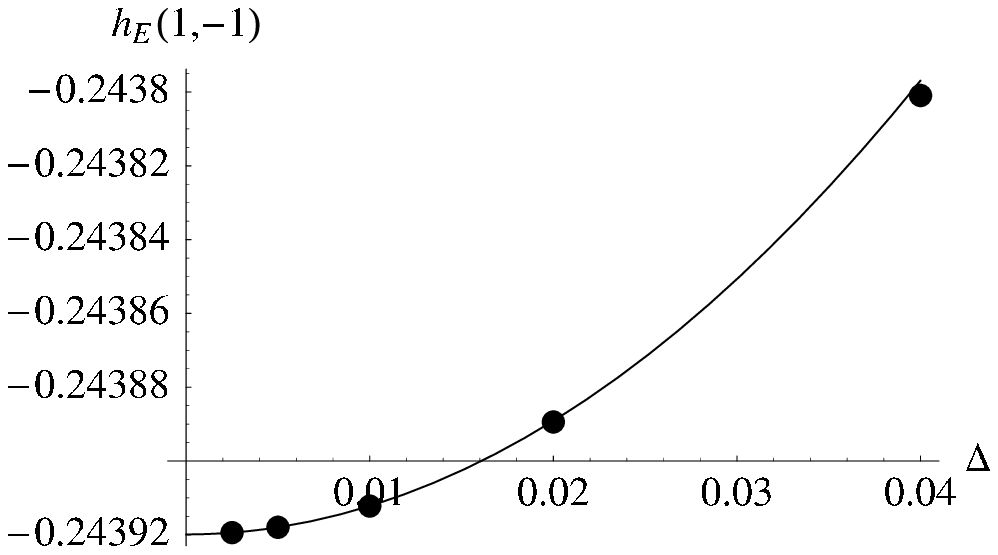,width=2in}
\hskip 0.1in b) \epsfig{file=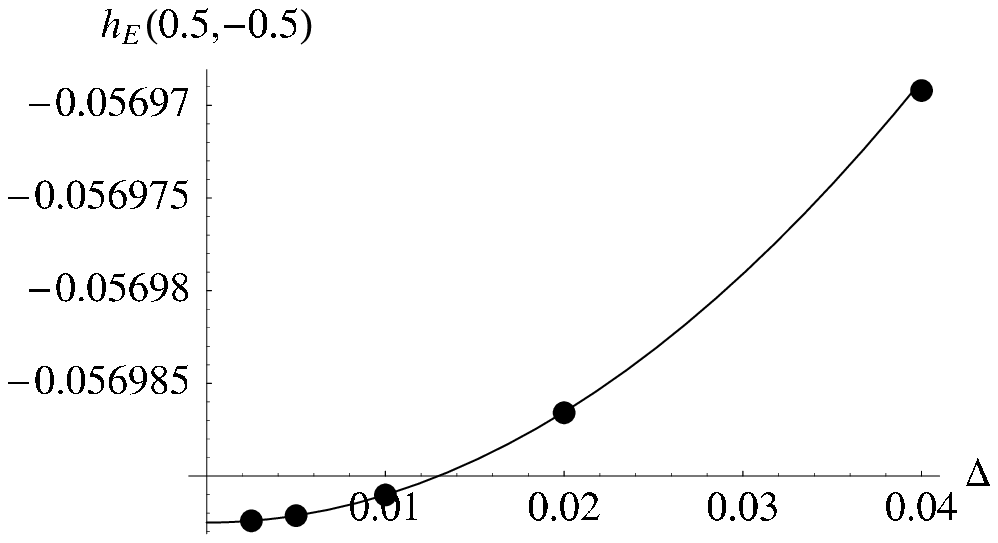,width=2in}
\hskip 0.1in c) \epsfig{file=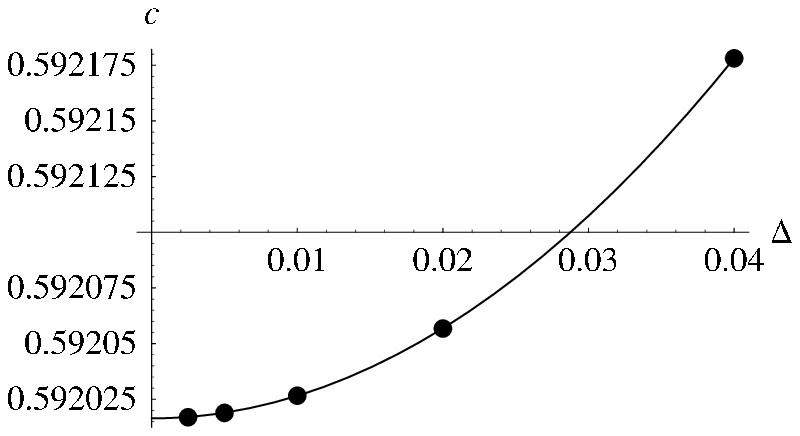, width=2in}
}
\caption{(a) $h_{\rm E}(1,-1)$ and (b) $h_{\rm E}(0.5,0.5)$ as functions of the number of lattice points, compared to a fit assuming second order scaling to the continuum, fitted using the highest two resolution points. Then (c) shows the value of $c$ such that $h(0,0)$ vanishes, the extrapolated continuum value of which is 0.592016.}
\label{continuum}
}

In both the complex and symplectic implementations various resolutions were used to compute the K\"ahler-Einstein metric, both to check convergence to the continuum and estimate error.

Taking the example of the symplectic implementation, we uniformly covered the coordinate square $0 \le x_1 \le 1, -1 \le x_2 \le 0$. Various lattice sizes were used, including $25\times25, 50\times50, 100\times100, 200\times200, 400\times400$. The calculations were performed on a desktop computer, with the lowest resolution $25\times25$ taking seconds to run, and the highest resolution $400\times400$ taking many hours. We estimate $h_{\rm E}$ as $h$ for sufficient flow time that the update of $h$ in a time step is of order the machine precision. Since second order differencing was used to implement the flow locally, we expect that any quantity measured, say $O$, should scale to the continuum as $O_{\mathrm{continuum}} + \ls^2 O_{\mathrm{correction}} + {\mathcal O}(\ls^3)$. Values of the relaxed function $h$ at different coordinate locations were used to check this scaling, and indeed give this consistent second order scaling behaviour. In Fig.~\ref{continuum} we give an example, plotting the value of the boundary point $h_{\rm E}(1,-1)$, and also the value of an interior point $h_{\rm E}(0.5,-0.5)$ where we note that in the flow, $h(0,0)$ is fixed to zero by the appropriate choice of $c$,
which is shown in Fig.~\ref{continuum}c. 
Using the two highest resolution points we fit the second order scaling behaviour above, and see a very good fit to the lower resolution points, with $h_{\rm E}(1,-1) = -0.2439 + 0.077 \ls^2$ and $h_{\rm E}(0.5,-0.5) = -0.0570 + 0.014 \ls^2$.  These fits indicate the error in the value of $h_{\rm E}$ at a point
is about $10^{-6}$ for the highest resolution $400 \times 400$ grid calculated in the symplectic implementation.

\FIGURE{
\centerline{ \psfig{figure=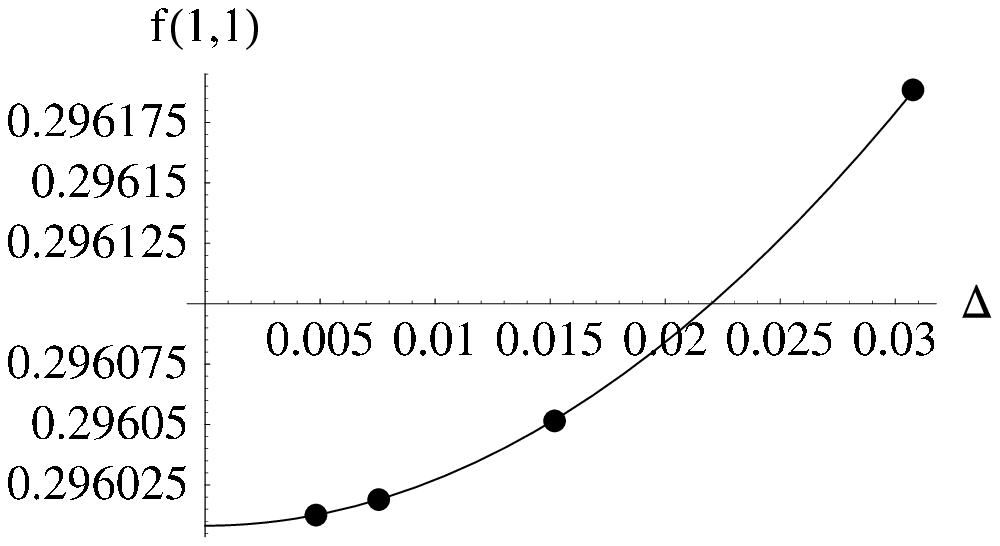, width=3in}}
\caption{A plot of the value
of the K\"ahler potential at the center of the hexagon as a function of $\ls$. The four data points correspond to our $40\times 40$, $80 \times 80$, $160\times 160$, 
and $250\times 250$ grids. The linear fit was made with the two data points with smallest $\ls$, giving $0.296008 + 0.1898 \ls^2$}
\label{linearplots}
}
Likewise in the complex coordinate implementation, various resolutions were computed, up to $250 \times 250$. The value of the potential at the center of the hexagon, $(\xi=1, \eta=1)$ is plotted in Fig.~\ref{linearplots} and we see that the convergence to the continuum value is quadratic in $\ls$, again consistent with the second order spatial finite differencing. This and other such tests suggest our best $250 \times 250$ grid in the complex case is accurate to about a part in $10^{5}$. 
\FIGURE{
\epsfig{file=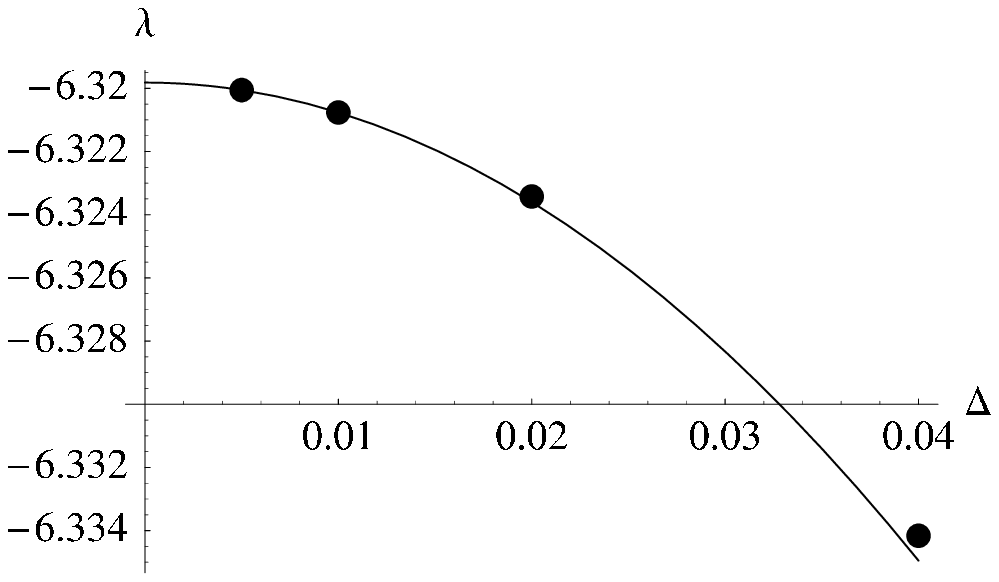,width=2.5in}
\caption{The eigenvalue as a function of number of lattice points for the lowest non-constant eigenfunction which transforms trivially under the hexagon and $U(1) \times U(1)$ symmetry. Again we see good agreement with a second order scaling fit function.}
\label{evalcont}
}
Note that in both the symplectic and complex coordinates, the value of $2\F + c$ at the center
of the hexagon was found to be $0.592016$.

The diffusion flow used to study the eigenfunctions of the scalar Laplacian was also differenced to second order accuracy. The flow was simulated using the same explicit method, with time step $\frac12\ls^2$, and for the same resolutions as above. Using the lowest non-constant eigenfunction which transforms trivially under the action of the $U(1)^2$ and $D_6$ isometries, we plot in Fig.~\ref{evalcont} the eigenvalue extracted for different resolutions, using the same initial data for the diffusion, and again fitting second order scaling to the two highest resolutions.  We see consistent second order scaling behaviour.

\section{$D_6$-invariant polynomials}
\label{app:poly}

In this section, we describe the set of polynomials
\be
{\mathcal P} = \sum_{n,m} b_{n,m} x_1^n x_2^m 
\ee
invariant under the dihedral group acting on the hexagon.

Our dihedral group is generated by two elements $R_1$ and $R_2$
(see (\ref{dihedral})).
%\be
%R_1 = \left(
%\begin{array}{rc}
%1 & 1 \\
%-1 & 0 
%\end{array}
%\right) \; \mbox{and} \; \;
%R_2 = 
%\left(
%\begin{array}{cc}
%0 & 1 \\
%1 & 0 
%\end{array}
%\right) \ .
%\ee
If we take a linear combination of the $x_i$, $v \cdot x$, then
$R_i (v \cdot x) = v \cdot R_i \cdot x = (R^t_i v) \cdot x$, 
i.e~the $R_i$ act on $v$ by their transpose.

Now, $R_2$ has eigenvalues $e^{\pi i/3}$ and  $e^{-\pi i/3}$.
A convenient basis of eigenvectors is
\begin{eqnarray*}
v_1 &=& (e^{\pi i/6} , e^{-\pi i/6}) \ , \\
v_2 &=& (e^{-\pi i/6} , e^{\pi i/6}) \ .
\end{eqnarray*}
The basis is convenient because a polynomial that is symmetric in 
$a_1 \equiv v_1\cdot x$ and $a_2 \equiv v_2 \cdot x$
is symmetric in $x_1$ and $x_2$. 

Consider the polynomial
\be
{\mathcal P}  = \sum_{n,m} c_{n,m} a_1^n a_2^m \ .
\ee
Symmetry under interchange of $a_1$ and $a_2$ requires $c_{n,m} = c_{m,n}$.
Moreover,
we have
\be
R_2 (a_1^n a_2^m) = \exp \left( \frac{\pi i}{3}(n-m) \right) a_1^n a_2^m
\ee
which implies that $n-m \equiv 0 \mod 6$.
We conclude that the most general polynomial invariant under the group
action can be decomposed into a sum of polynomials of the form
$a_1^n a_2^m + a_1^m a_2^n$ where $n-m \equiv 0 \mod 6$.

For example, two important invariant polynomials are
\begin{eqnarray*}
a_1 a_2 &=& x_1^2 + x_1 x_2 + x_2^2 \equiv U \ , \\
\frac{1}{27} (a_1^3+a_2^3)^2 &=& x_1^2 x_2^2 (x_1+x_2)^2 \equiv V \ .
\end{eqnarray*}

We now argue that any invariant polynomial can be decomposed into
sums and products of $U$ and $V$.
Assume $a_1^n a_2^m + a_1^m a_2^n$ is left invariant by $R_2$
and assume $m$ is the minimum of $m$ and $n$.
Then 
\[
a_1^n a_2^m + a_1^m a_2^n = (a_1 a_2)^{m} (a_1^{n-m} + a_2^{n-m})
= U^m (a_1^{n-m}+a_2^{n-m})
\]
One can prove inductively that 
$a_1^{n-m}+a_2^{n-m}$ can be written in terms of $U$ and $V$ 
whenever $(n-m)$ is a multiple of 6. So $U$ and $V$ 
generate the polynomials invariant under $R_1$ and $R_2$.

\section{Convergence and error estimates of the constrained optimization method}
\label{app:convfit}

In this appendix we give convergence results for the smooth part of the symplectic potential, $h(x_1, x_2)$ determined in Section \ref{sec:leastsquares}. At all points in the hexagon domain the value of $h$ was observed to converge quickly with increasing numbers of terms taken in the series expansion at the origin. In Fig.~\ref{contfit} we plot the value of $h$ at 3 points. Note that one of these points lies on the boundary of the hexagon. All other points checked, both in the hexagon interior and on the boundary, gave qualitatively similar convergence. In the figure we also compare the data with the extrapolated continuum results for the Ricci flow given in the previous appendix (suitably adjusting for the different value of $c$). We see excellent agreement. We may also estimate that the eighteenth order results and the estimated infinite order result differ at the 1 part in $10^6$ level.

\FIGURE{
\centerline{a) \epsfig{file=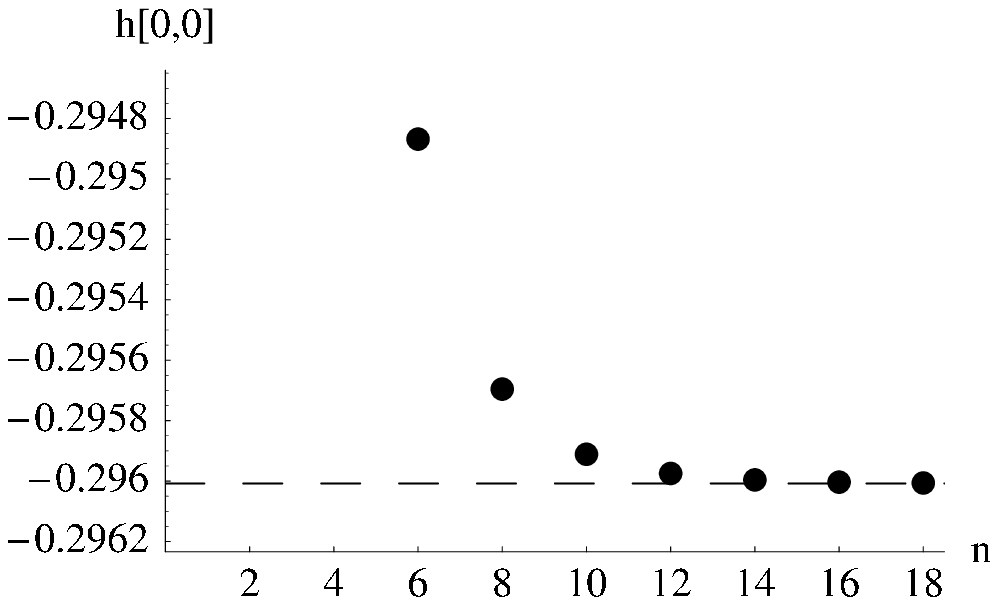,width=2in}
\hskip 0.1in b) \epsfig{file=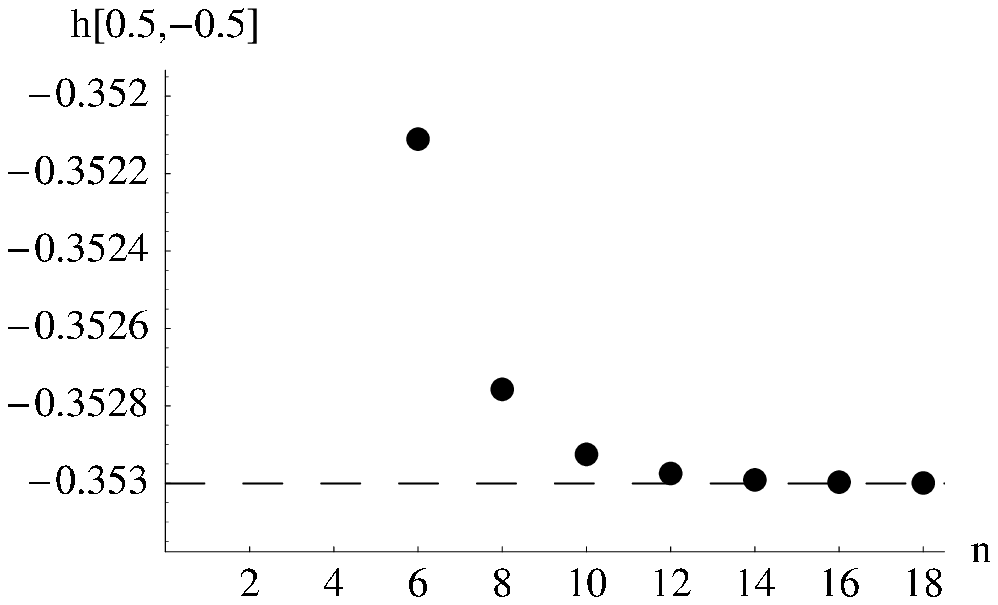,width=2in}
\hskip 0.1in c) \epsfig{file=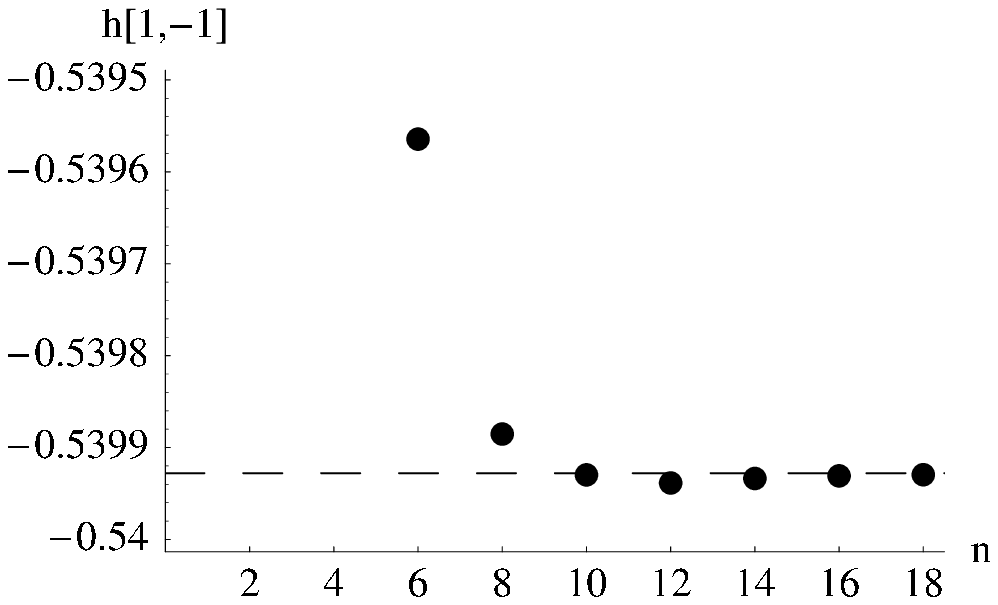, width=2in}
}
\caption{(a) $h(0,0)$, (a) $h(0.5,-0.5)$ and (b) $h(1,-1)$ as functions of the number terms in the expansion. The dashed lines show the extrapolated continuum from the Ricci flow results.}
\label{contfit}
}

\end{appendix}

\end{document}